\documentclass[twocolumn]{aastex63}

\usepackage{todonotes}
\usepackage{bm}
\usepackage{graphicx}	
\usepackage{amsmath}	
\usepackage{amssymb}	
\usepackage{longtable}
\usepackage{subfigure}
\usepackage{rotating}
\usepackage{float}
\usepackage{natbib}
\usepackage{xspace}
\usepackage{epstopdf}
\usepackage[]{url}
\usepackage{enumitem}
\usepackage[normalem]{ulem}
\usepackage{xcolor}

\newcommand{\Msun}{M$_{\odot}$\xspace}
\newcommand{\Mvir}{M$_{\rm vir}$\xspace}
\newcommand{\Rvir}{R$_{\rm vir}$\xspace}
\newcommand{\kms}{km s$^{-1}$\xspace}

\received{\today}
\submitjournal{ApJ}

\shorttitle{Evidence for a Massive M31}
\shortauthors{Patel \& Mandel}
\graphicspath{{./}{figures/}}

\defcitealias{patel17b}{P17}
\defcitealias{patel18a}{P18}

\begin{document}

\title{Evidence for a Massive Andromeda Galaxy Using Satellite Galaxy Proper Motions}

\correspondingauthor{Ekta Patel}
\email{ektapatel@berkeley.edu}

\author[0000-0002-9820-1219]{Ekta Patel}
\affiliation{Department of Astronomy, University of California, Berkeley, 501 Campbell Hall, Berkeley, CA, 94720, USA} 

\author[0000-0001-9846-4417]{Kaisey S. Mandel}
\affiliation{Institute of Astronomy and Kavli Institute for Cosmology, Madingley Road, Cambridge CB3 0HA, UK}
\affiliation{Statistical Laboratory, DPMMS, University of Cambridge, Wilberforce Road, Cambridge CB3 0WB, UK}
\affiliation{The Alan Turing Institute, Euston Road, London NW1 2DB, UK}

\begin{abstract}
We present new mass estimates for Andromeda (M31) using the orbital angular momenta of four satellite galaxies (M33, NGC~185, NGC~147, IC~10) derived from existing proper motions, distances, and line-of-sight velocities. We infer two masses for M31: $M_{\rm vir}= 2.85^{+1.47}_{-0.77}\times10^{12}\, M_{\odot}$ using satellite galaxy phase space information derived with \textit{HST}-based M31 proper motions and $M_{\rm vir}=3.02^{+1.30}_{-0.69}\times10^{12}\, M_{\odot}$ using phase space information derived with the weighted average of \textit{HST+Gaia}-based M31 proper motions. The precision of our new M31 mass estimates (23-50\%) improves by a factor of two compared to previous mass estimates using a similar methodology with just one satellite galaxy and places our results amongst the highest precision M31 estimates in recent literature. Furthermore, our results are consistent with recently revised estimates for the total mass of the Local Group (LG), with the stellar mass--halo mass relation, and with observed kinematic data for both M31 and its entire population of satellites. An M31 mass $> 2.5 \times 10^{12}\, M_{\odot}$ could have major implications for our understanding of LG dynamics, M31's merger and accretion history, and our understanding of LG galaxies in a cosmological context.

\end{abstract}



\section{Introduction} 
\label{sec:intro}
Pinning down the total masses of the Milky Way (MW) and Andromeda (M31) is vital to almost all aspects of understanding the formation and evolution of the Local Group (LG). Nearly all galaxy parameters are directly correlated to the total mass of a galaxy, a majority of which resides in the dark matter halo. Therefore, constraining halo mass is also key to revealing clues about the nature of dark matter itself.  

While recent works have made significant progress towards constraining the total mass of the MW using methods that rely on measured 6D phase space information (3D position + 3D velocity) for various stellar substructures, including satellite galaxies, globular clusters, stellar streams, and halo stars \citep[e.g.,][]{busha11,bk13,gonzalez13, penarrubia14, cautun14, gibbons14, penarrubia16,eadie17, deason21}, the lack of equivalent information for substructures around M31 has posed a number of challenges in advancing our understanding of M31. One key piece of missing information has been a precise mass estimate that reconciles the latest picture of M31's merger history, accretion history, and observed properties.

Nevertheless, a variety of methods have been previously utilized to estimate the mass of M31 in the absence of 6D phase space information. However, systematic differences in the assumptions, methods, and data have still led to a large scatter ranging from $0.5-3.5 \times 10^{12} \, M_{\odot}$. Examples of such assumptions include the use of a steady-state halo and spherical halo geometry, which may no longer be accurate in light of recent studies that have shown that M31 has likely had an eventful recent past \citep[e.g.,][]{lewis23, mackey19, dsouza18, hammer18, mcconnachie18, mcconnachie09, fardal06, dey22}. Other assumptions, such as those requiring constant velocity anisotropy, can either over- or underestimate the uncertainty in the mass of M31, demonstrating the need for less biased mass estimation techniques.

Another commonly used technique that yields high uncertainty in virial mass\footnote{``Virial" quantities refer to quantities calculated following the definitions provided in \citet{brynorman98}.} (or equivalent total halo mass definitions, such as $M_{200}$) is the process of extrapolating between enclosed masses to virial masses. The former have less associated uncertainty, however, virial masses are often more useful for placing a galaxy in a cosmological context. Enclosed masses have been reported for methods including those that use galaxy rotation curves \citep[e.g.,][]{chemin09,sofue15}, distribution functions \citep[e.g.,][]{evans00, evans_wilkinson00}, the Jeans equation \citep[e.g.,][]{watkins10}, stellar streams \citep{ibata04, fardal06, fardal13, dey22}, globular clusters \citep[e.g,][]{perrett02, lee08, galleti06, veljanoski13}, and sometimes satellite galaxies \citep[e.g.,][]{hayashi_chiba14}. Strong assumptions combined with extrapolation techniques can then further reduce both the accuracy and precision of mass estimation techniques. Therefore, in this work, we turn to high-precision astrometric data and cosmological simulations to devise a technique that statistically constrains host galaxy virial mass, eliminating these extra sources of uncertainty. 

While 6D phase space information is available for almost all of the MW's satellite galaxies \citep[][]{gaiadr2b,simon18, fritz18, mcconnachie20a, mcconnachie20b, li21, battaglia22, pace22}, the same information is only available for four of M31's satellite galaxies (M33, NGC~185, NGC~147, IC~10) owing to their distance \citep{brunthaler05,brunthaler07,sohn20}. The proper motion (PM) of M31 itself was also only measured a decade ago for the first time \citep{sohn12, vdm12ii}.

In previous work, we demonstrated that combining cosmological simulations with high-precision data for satellite galaxies is a powerful technique to constrain host galaxy virial mass, building on the statistical methods of \citet{busha11} and \citet{gonzalez13}. In \citet[][hereafter P17]{patel17b} we used a Bayesian framework to estimate the mass of M31 and the MW using measurements of the following observed properties of the Large Magellanic Cloud (LMC) and M33: distance relative to the MW or M31 ($r^{\rm obs}$), velocity relative to the MW or M31 ($v^{\rm obs}_{\rm tot}$), maximum circular velocity ($v^{\rm obs}_{\rm max}$), and orbital angular momentum ($j^{\rm obs}$). This approach included two sub-methods, the first using the combination of instantaneous properties, namely position and velocity, and the second focusing on dynamical properties, particularly orbital angular momentum. We found the best estimate of M31's mass from each sub-method to be \Mvir$=1.44^{+1.26}_{-0.69}\times10^{12}\, M_{\odot}$ (instantaneous method) and \Mvir$=1.37^{+1.39}_{-0.75}\times10^{12}\, M_{\odot}$ (momentum method). Furthermore, we concluded that angular momentum is a much more reliable tracer of host galaxy halo mass, as it is robust against bias introduced by different host-satellite orbital configurations and whether a satellite is bound to its host or not.

Extending the analysis of \citetalias{patel17b}, in \citet[][hereafter P18]{patel18a} eight satellite galaxies were used to estimate the mass of the MW to a precision of $\sim$30\%, significantly improving on the precision achievable with only one satellite. However, given the low number of halos and subhalos in simulations that broadly represent the properties of MW satellites, a post-processing step was introduced to statistically combine the posterior distributions of MW virial mass inferred with each satellite. Thus, in practice, our ``ensemble" results, which aim to leverage the 6D phase space of eight MW satellite galaxies simultaneously are still an approximation of the MW's mass, especially since the mass resulting from analyzing individual satellites independently exhibits a scatter of a factor of three.

Here, we improve on the work of \citetalias{patel17b} by including the phase space information for three additional M31 satellites. Given the smaller satellite sample size compared to \citetalias{patel18a}, we are also able to improve our statistical methodology. In particular, we can relax the statistical approximation previously needed to combine posterior distributions corresponding to individual satellites and instead compute joint likelihoods with four satellites simultaneously. This is expected to yield the most precise M31 mass to date, allowing all properties of M31's halo, including its shape, to act as free parameters and requiring no assumptions about whether satellites are bound to M31. 

This paper is organized as follows. In Section \ref{sec:data} we describe both the observational data sets and the simulations used in this analysis. Section \ref{sec:methods} briefly outlines the statistical methods from previous work and the modifications required for the M31 system. In Section \ref{sec:result} we provide results for the estimated mass of M31, and Section \ref{sec:discussion} places these results in the context of recent literature and cosmological predictions. Finally, we conclude and summarize in Section \ref{sec:summary}.

\section{Simulated and Observed Galaxy Properties}
\label{sec:data}
In this section, we discuss the observed data for M31 and the four M31 satellite galaxies of interest (NGC~147, NGC~185, M33, IC~10). We also discuss the IllustrisTNG-Dark simulations, the suite of dark matter only simulations used in combination with the observed data to statistically estimate the mass of M31.

\subsection{Observed Satellite Properties}
\subsubsection{Distances and Radial Velocities}
We use distances from a novel homogeneous compilation \citep{savino22, nagarajan22} that have been derived as part of the Cycle 27 \textit{HST} Treasury survey of the M31 system (GO-15902, PI: D. Weisz) for all galaxies except IC~10. The distances to M31 and to the satellite galaxies have been measured from \textit{HST} time-series of RR Lyrae variable stars, using a reddening-independent calibration from the models of \citet{marconi15}, which have been empirically re-calibrated to ensure consistency with the \textit{Gaia} eDR3 astrometric reference frame. Measuring a new distance to IC~10 was not possible due to high extinction, so we adopt the \citet{mcquinn17} distance based on the tip of the red giant branch. All distance moduli and heliocentric line-of-sight (LOS) radial velocities are listed in Table \ref{tab:distmod_vlos}.

\begin{table*}
\centering
    \begin{tabular}{|c|c|c|c|c|}\hline
     Galaxy    &  $(m-M)_0$  & $v_{ \rm LOS}$  & $\mu_{\alpha*},\,  \mu_\delta$ & Refs. \\ 
     & [mag] & [km s$^{-1}$] & [$\mu$as yr$^{-1}$]  &\\ \hline
        M31 (\textit{HST}+sats) & 24.45$\pm$0.06 & -301.0 & 45$\pm$13, -32$\pm$12 & 1,3,4\\ \hline
        
        M31 (\textit{HST}+\textit{Gaia}) & 24.45$\pm$0.06 & -301.0 & 49$\pm$11, -38$\pm$11 & 1,4,8 \\ \hline
        M33 & 24.67 $\pm$0.06 & -179.2 & 23$\pm$7, 8$\pm$9 & 1,5,9 \\ \hline
        
        NGC~185 & 24.06$\pm$0.06 & -203.8 & 24$\pm$14, 6$\pm$15 &  1,6,11 \\ \hline
        
        NGC~147 & 24.33$\pm$0.06  & -193.1 & 23$\pm$14, 38$\pm$15 &  1,6,11 \\ \hline
        
        IC~10 &  24.43$\pm$0.03 & -348.0 & 39$\pm$9, 31$\pm$8 & 2,7,10  \\ \hline
    \end{tabular}
    \caption{Distance moduli, LOS radial velocities, and proper motions for M31 and the four satellite galaxies used in this work. References are labeled as follows: (1) \citet{savino22}; (2) \citet{mcquinn17}; (3) \citet{vdm12ii}; (4) \cite{slipher13}; (5) \citet{corbelli97}; (6) \citet{geha10}; (7) \citet{huchra99}; (8) \citet{vdm19}; (9) \citet{brunthaler05}; (10) \citet{brunthaler07}; (11) \citet{sohn20}.}
    \label{tab:distmod_vlos}
\end{table*}

\subsubsection{M31}
This work relies on the properties of satellite galaxies with respect to M31, thus we first need to establish the M31 properties that will be used to subsequently derive observed satellite galaxy properties. We use the M31 distance derived by \citet{savino22} which gives $D_{M31}=776.2^{+22}_{-21}$ kpc. The LOS velocity, $ v_{\rm LOS}=-301$ \kms, comes from \citet{slipher13}. The last necessary component to convert the properties of satellite galaxies to an M31-centric reference frame is the Galactocentric motion of M31, which relies on a measured PM. This velocity acts as a zero-point of the satellites' motion. 

The first direct PM measurement for M31 was taken with the {\em Hubble Space Telescope} (\textit{HST}) in 2012 \citep{sohn12,vdm12ii}. Multiple additional estimates (both direct and indirect) for M31's PM have also been reported using satellite galaxies and stellar population data measured with both \textit{HST} and \textit{Gaia} \cite[e.g.,][]{vdmG08, salomon16, vdm19, salomon21}. As in \cite{sohn20}, we adopt two M31 PM measurements, those reported in \citet[][referred to as \textit{HST}+sats]{vdm12ii} and \citet[][referred to as \textit{HST}+\textit{Gaia}]{vdm19}, which give tangential velocity zero-points of $V_{\rm tan, HST+sats}=17 \rm \,km \, s^{-1}$ (with a $1\sigma$ confidence region of $V_{\rm tan, HST+sats} \le 34.3$ km s$^{-1}$) and $V_{\rm tan, HST+GaiaDR2}=57^{+35}_{-31} \rm \, km\, s^{-1}$, respectively (see Table \ref{tab:obsdata}). 

We do not consider the measurements reported in \citet{salomon21} using data from \textit{Gaia} eDR3 as their PMs measured with blue young main sequence stars are consistent with and as accurate as the weighted average between \textit{HST} measurements and indirect estimates from the LOS velocities of satellite galaxies \citep[our \textit{HST}+sats data;][see Fig. 6 in \citet{salomon21}]{sohn12,vdm12ii}. Throughout this analysis, we will present results using observational satellite properties derived from both sets of M31 tangential velocity zero points. Table \ref{tab:obsdata} lists the relevant observational properties for the four satellite galaxies used in this study. In the top half, the positions, velocities, and angular momenta are derived using the M31 \textit{HST}+sats tangential velocity zero-point, and in the bottom half, the \textit{HST}+\textit{Gaia} tangential velocity zero-point is used. 

\subsubsection{M33}
Position, velocity, and orbital angular momentum for M33 using the M31 \textit{HST}+sats zero-point are adopted from \citetalias{patel17b} (see Table 1 and references therein) and are listed in the first row of Table \ref{tab:obsdata}. In short, M33's PM was first measured using the {\em Very Long Baseline Array} (\textit{VLBA}) by \cite{brunthaler05}. 
M33's PM was also independently measured again in \citet{vdm19} using data from \textit{Gaia} DR2. In this analysis, we adopt the weighted average of the \textit{VLBA}+\textit{Gaia} DR2 PMs from \citet{vdm19} whenever the M31 \textit{HST}+\textit{Gaia} DR2 tangential velocity zero-point is used (bottom half of Table \ref{tab:obsdata}). The most substantial changes between the \textit{HST}+sats and \textit{HST}+\text{Gaia} data sets are that M33's 3D velocity vector relative to M31 increased by 55 km s$^{-1}$ and the 3D position vector relative to M31 increased by $\sim$ 20 kpc. 

Our framework relies on the maximum circular velocity of only the dark matter halo of a satellite's rotation curve, $v_{\rm max}^{\rm obs}$. M33's total rotation curve was measured out to $\sim$15 kpc by \cite{corbellisalucci} reaching a maximum velocity of $\approx 130$ km s$^{-1}$. For this value, we use the peak halo velocity from \citet[][90 km s$^{-1}$]{vdm12iii} for M33, which is determined by reconstructing a model rotation curve to match the observed data. The M33 \textit{VLBA} PMs are used to determine the values listed in the top half of Table \ref{tab:obsdata}.

We calculate the uncertainties on position, velocity, and orbital angular momentum using Monte Carlo draws from the $4\sigma$ error space on the measured LOS velocity, distance modulus, and PM measurements \citep[see][]{patel18a, vdm02}. The values and associated uncertainties listed in Table \ref{tab:obsdata} represent the mean and standard deviation for each quantity using 10,000 position and velocity vectors resulting from the Monte Carlo sampling. We assign an uncertainty of 10 km s$^{-1}$ to $v_{\rm max}^{\rm obs}$ for all satellites since this is equivalent to reported uncertainties in rotation curves for galaxies at these distances. 

Note that the new distance for M33 \citep[226 kpc in this work vs. 203 kpc in \citetalias{patel17b};][]{savino22} increases the angular momentum from 27656$\pm$8219 kpc km s$^{-1}$ \citepalias{patel17b} to 37158$\pm$8011 kpc km s$^{-1}$ (this work), an approximate increase of 30\%. We have also improved our methodology for drawing Monte Carlo samples since \citetalias{patel17b}, however, these methodological changes only affects the tails of M33's $j$ distribution. Despite changes in adopted $j$, M31's mass resulting from the properties of M33 are still consistent within the uncertainties of the \citetalias{patel17b} value, and conclusions from \cite{patel17a} that M33 is likely to be on a first infall orbit still hold. See also Appendix \ref{app:tng_results}.

\subsubsection{NGC~147 \& NGC~185}
\label{subsubsec:ngcs}
 The first PMs of NGC~147 and NGC~185 were recently measured using \textit{HST} \citep{sohn20}. To determine the appropriate maximum circular velocity values of these two dwarf elliptical galaxies, we first use the abundance matching relation from \citet{moster13}\footnote{These abundance matched masses are consistent with those from \citet{gk17a}, despite the well-known discrepancies for different SMHM relations in the regime of low-mass galaxies.} to find the infall halo mass for NGC~147 and NGC~185 using the stellar masses reported in \citet{mcconnachie18}. 
From this relation, we determine the following infall halo masses: $5\times10^{10}\, M_{\odot}$ (NGC~147) and $4.5\times10^{10}\, M_{\odot}$ (NGC~185). Using these halo masses, we construct individual NFW halo profiles to best fit the dynamical masses reported in \citet{mcconnachie12}. Dynamical mass relies on the half-light radius and the Wolf estimator \citep{wolf10} to constrain the mass within the half-light radius. We varied the concentration of each NFW profile until the enclosed mass at the half-light radius matched the dynamical mass. The best-fit NFW halo profiles result in maximum circular velocities of 61 km s$^{-1}$ (NGC~147) and 73 km s$^{-1}$ (NGC~185).

\subsubsection{IC~10}
\label{subsubsec:ic10}
We use the PM of IC~10, the furthest satellite galaxy considered in this work, as measured with the \textit{VLBA} \citep{brunthaler07}. To determine the maximum circular velocity of IC~10's dark matter halo, we adopt the following properties from Table 2 of \citet{oh15}: $R_{max}$, $M_{dyn}(R_{max})$, and $M_{200}$. We first convert $M_{200}$ to virial units, which yields a virial mass of \Mvir$=1.9\times10^{10} \, M_{\odot}$. Then, we subtract the stellar and gaseous masses from \Mvir and use this mass to construct an NFW profile. We vary the concentration of the NFW profile, until the NFW profile's enclosed mass at $R_{max}$ is equivalent to $M_{dyn}(R_{max})$. The best-fitting NFW profile results in a maximum circular velocity of 48 km s$^{-1}$ for IC~10's halo.

\begin{table}[]
    \centering
    \begin{tabular}{|c|c|c|c|c|}\hline
     Galaxy    &  $r^{\rm obs}$  & $v^{\rm obs}_{\rm max}$  & $v^{\rm obs}_{\rm tot}$ &  $j^{\rm obs}$  \\ 
     & [kpc] & [km s$^{-1}$] & [km s$^{-1}$] & [kpc km s$^{-1}$] \\ \hline
     \multicolumn{5}{|c|}{M31 HST+sats $v_{\rm tan}$ zero-point}  \\ \hline
        M33$^a$ & 226$\pm$13 & 90$\pm$10 & 202$\pm$40 &  38253$\pm$8010$^c$ \\
        NGC~185 & 155$\pm$25 & 73$\pm$10 & 127$\pm$31& 18113$\pm$7366\\
        NGC~147 & 107$\pm$12 & 61$\pm$10 &205$\pm$47 & 18082$\pm$6091 \\
        IC~10 & 247$\pm$24 & 48$\pm$10 & 264$\pm$44 & 56687$\pm$11063 \\ \hline
        \multicolumn{5}{|c|}{M31 HST+Gaia DR2 $v_{\rm tan}$ zero-point}  \\ \hline
        M33$^b$ & 226$\pm$13 & 90$\pm$10 & 257$\pm$50 & 41502$\pm9886$ \\
        NGC~185 & 155$\pm$25 & 73$\pm$10 & 194$\pm$48 & 29832$\pm$9740\\ 
        NGC~147 & 107$\pm$12 & 61$\pm$10 & 277$\pm$58 & 22219$\pm$8561 \\
        IC~10 & 247$\pm$24 & 48$\pm$10 & 339$\pm$51 &  66715$\pm$13045\\ \hline
    \end{tabular}
    \caption{The adopted observed data for all four M31 satellite galaxies with measured PMs and radial velocities. Properties include distance relative to M31 ($r^{\rm obs}$), velocity relative to M31 ($v^{\rm obs}_{\rm tot}$), maximum circular velocity ($v^{\rm obs}_{\rm max}$), and orbital angular momentum ($j^{\rm obs}$). a: M33's adopted PM is from the \textit{VLBA} \citep{brunthaler05}. b: M33's PM is the weighted average between \textit{VLBA} and \textit{Gaia} DR2 \citep{vdm19}. c: In \citetalias{patel17b}, we adopted $r^{\rm obs}$ = 203 kpc and thus $j=27656\pm8219\rm \, kpc\ km\,s^{-1}$ for M33 but since we have updated M33's distance to $r^{\rm obs}$ = 226 kpc in this work, $j$ for M33 has kpc also increased.}
    \label{tab:obsdata}
\end{table}

\subsection{IllustrisTNG-Dark}
We use halo catalogs from the IllustrisTNG project \citep{nelson18, pillepich18, naiman18, springel18, marinacci18, nelson19} to choose a broad range of host halos and their corresponding satellites as our prior sample. IllustrisTNG is a suite of hydrodynamical+N-body cosmological simulations. For this work, we specifically focus on IllustrisTNG100-1-Dark (hereafter IllustrisTNG-Dark), which follows the evolution of 1820$^3$ dark matter particles from $z\approx 127$ to $z=0$. Each dark matter particle has a mass of $m_{DM} =6\times10^{6}\,M_{\odot}/h$. The following cosmological parameters are adopted for consistency with the results from \citet{planck15}: $\Omega_{\Lambda,0}=0.6911$, $\Omega_{m,0}=0.3089$, $\Omega_{b,0}=0.0486$, $\sigma_8=0.8159$, $n_s=0.9667$, and $h=0.6774$. 

As discussed in \citetalias{patel18a}, we focus on a dark matter-only simulation since it yields the largest prior sample (i.e., fewer satellites are disrupted or inhibited from forming due to baryonic effects). However, full hydrodynamics still yields a consistent answer for MW halo masses as compared to the dark matter-only simulations (see \citetalias{patel18a}, Fig. 6).

\section{Statistical Methods}
\label{sec:methods}
In \citetalias{patel18a}, we used eight classical MW satellites to estimate the mass of the MW but we were limited by the number of representative subhalos and corresponding host halos in Illustris-1-Dark (hereafter Illustris-Dark). We found that halos in Illustris-Dark at low redshift typically host between two and five subhalos with properties broadly representative of the MW satellites. Therefore, likelihood functions were evaluated per individual satellite using the same prior sample and the results were combined using a statistical approximation that removed the multiplicity of using the same prior for each satellite. 

Since 6D phase space information is only available for four M31 satellites, it is possible to take a more rigorous approach to estimate the mass of M31. In short, there are enough halos hosting four subhalos representative of these M31 satellites, and thus we do not need the additional approximation necessary to combine the results for individual satellites as was needed in \citetalias{patel18a}. It does, however, require building a more strategic prior sample and a modification of the likelihood functions previously implemented in \citetalias{patel18a}.

\subsection{Prior Sample Selection}
\label{subsec:prior}
For host halos and subhalos in the prior sample, the physical properties of interest are $\bm{\Theta} = [\bm{X}, m]$ where $m \equiv \log_{10} M_{\rm vir}$ and \Mvir is the virial mass\footnote{We refer to virial mass and virial radius throughout this work. In IllustrisTNG-Dark, these represent the virial mass and radius of FoF groups as identified by \texttt{SUBFIND}. \texttt{SUBFIND} definitions follow those of \citet{brynorman98}. Adopting the IllustrisTNG-Dark cosmology, this corresponds to $\Delta_{\rm vir}=330$.} of the host halo of any given subhalo. We define
\begin{equation}
\bm{X} = \{ \bm{x}_1, \ldots, \bm{x}_{N_{\rm sat}} \}
\end{equation}
as the collection of $N_{\rm sat}$ subhalo properties and $\bm x_s = [v_{{\rm max},s}, r_s, v_{{\rm tot},s}, j_s]$ are the latent, observable parameters of subhalo $s$. The observational data in Table \ref{tab:obsdata}, denoted as $\bm{d}_s = [v_{{\rm max},s}^{\rm obs}, r_s^{\rm obs}, v_{{\rm tot},s}^{\rm obs}, j_s^{\rm obs}]$, are measurements of the parameters $\bm{x}_s$ of the M31 satellites such that if the errors on the measurements of the parameters $\bm{x}_s$ were zero, then $\bm{d}_s = \bm{x}_s$. We define $\bm{D} = \{ \bm{d}_1, \ldots, \bm{d}_{N_{\rm sat}} \}$ as the collection of measurements of the $N_{\rm sat}$ subhalo properties.

To build a new prior sample, representing draws from $P(\bm{\Theta})$, that will simultaneously constrain the mass of M31 using the four satellites of interest, we first select all simulated halos where the host halo has a minimal virial mass $\geq10^{10}\, M_{\odot}$ at $z\approx0$ and $v_{\rm max} < 250$ \kms following the observed rotation curve \citep{corbelli10}. Recall that we select halos from snapshots 80-99 ($z=0-0.26$) to increase the total number of samples in the prior. While there may be repeated halo systems from snapshot to snapshot that satisfy the following criteria, we treat these halos as independent draws. Of these systems, those with four or more subhalos satisfying the following criteria $(\bm C)$ are considered:

\begin{itemize}\setlength{\itemindent}{-2em}
    \item[] $C_1$: the subhalo $v_{\rm max} = 35-100 \text{ km s}^{-1}$ at $z\approx0$
    \item[] $C_2$: the subhalo resides within 0.3$R_{\rm vir}$--$R_{\rm vir}$ at $z\approx0$
    \item[] $C_3$: the minimal subhalo mass is $\geq 5\times10^9\, M_{\odot}$ at $z\approx0$
    \item[] $C_4$: the subhalo is among the 10 most massive subhalos in the host halo group (excluding the primary)
\end{itemize}

 \noindent Previously, we required all subhalos to reside within the virial radius of their host halo, however, all four satellites are in the outer halo of M31 (two are at $> 100$ kpc and two are at $>$ 200 kpc). This particular subset of M31 satellites is not representative of the radial distribution of all M31 satellites where 17/35 or nearly 50\% (including NGC~147 and NGC~185) are located at 100-200 kpc, thus we modified this criterion to only include subhalos located outside of 0.3 \Rvir. Outer satellites are known to be the best tracers of the underlying host potential, therefore, this modification is advantageous to constraining the virial mass of M31. 

The prior sample purposefully only contains subhalos with $v_{\rm max} =35-100 \, \text{km s}^{-1}$ where 35 km s$^{-1}$ ensures that analogs of IC~10 are included but also that the typical $v_{\rm max}$ values where Too Big To Fail is most prominent around the MW and M31 ($\lesssim 40 \, \rm km \, s^{-1}$ in dark matter only simulations) are not considered \citep{bk11b, tollerud14}. Avoiding the TBTF challenge also supports our choice of using the dark matter-only version of IllustrisTNG-Dark. 

\begin{figure*}
    \centering
    \includegraphics[scale=0.42]{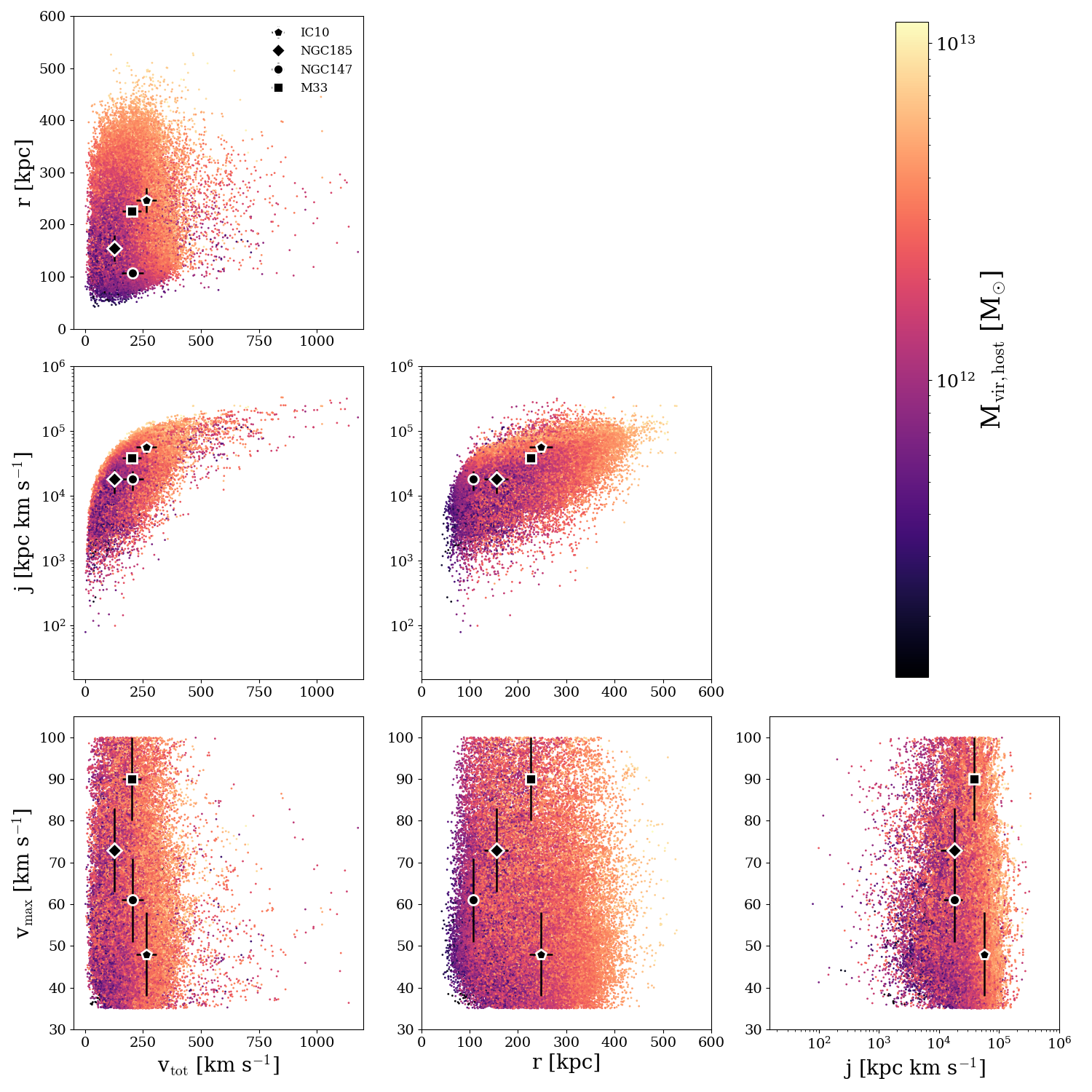}
    \caption{The distribution of subhalo properties for all host-satellite systems in the prior distribution, $P(m, \bm{x})$. Each panel shows a pair of properties in $\bm x$ and points are colored by the host halo virial mass. Observed satellite properties and their uncertainties, $\bm d$, with respect to the \textit{HST}+sats M31 $v_{\rm tan}$ zero-point are over-plotted in black markers to illustrate that the observed satellite properties and their measurement uncertainties are encompassed by the draws from the prior (see Section \ref{subsec:prior}). }
    \label{fig:prior}
\end{figure*}

These criteria yield 71,371 subhalos across 14,075 halo systems. Figure \ref{fig:prior} shows the distribution of latent subhalo properties, $\bm{x}$, for the IllustrisTNG-Dark subhalos in the prior sample compared to the observed properties of the four M31 satellites, $\bm{d}$. Each panel indicates a pair of two parameters in $\bm{x}$. The observed properties are adequately encompassed by the properties of the prior sample, indicating that this is an appropriate selection of subhalos. Qualitatively, Figure \ref{fig:prior} indicates that the observed properties are most similar to those systems with \Mvir $>10^{12}\, M_{\odot}$.

Note that while some systems have more than four subhalos satisfying these criteria, we only use the four subhalos with the highest $v_{\rm max}$ values in the following analysis since phase space information is only available for four M31 satellites. 

Since we consider four satellites with different observed properties, $\bm d$, we rank order the subhalos encompassed in the prior to ensure that the observed satellites are matched to the appropriate simulated counterpart. For each host halo in the prior, subhalos are ranked from highest to lowest $v_{\rm max}$, where $s=1$ represents the subhalo with the highest $v_{\rm max}$. The properties of the first subhalo in each system are statistically compared to the properties of M33 and so on, such that $s=1,2,..., N_{\rm sat}$ maps to M33, NGC~185, NGC~147, IC~10 and $N_{\rm sat}=4$, the total number of satellites. 

Therefore, though the draws from the prior defining our sample correspond to 71,371 subhalos, only 56,300 draws are used since just the first four subhalos in each halo system are considered. These subhalos/halos will be treated as draws from the underlying prior distribution.

\subsection{Likelihood Function}
\label{subsec:likelihoods}
To estimate the mass of M31, we use a subset of the parameters $\bm{x}$ in the following likelihood function, which is modified from those used in \citetalias{patel18a} to include any number of satellites, $N_{\rm sat}$. All observed satellite properties are assumed to have Gaussian errors. 

In \citetalias{patel17b} and \citetalias{patel18a}, we showed the advantages of the momentum method over the instantaneous method, which uses instantaneous properties like position and velocity, and therefore we only include results from the momentum method throughout the rest of this work.

For the angular momentum method, $\bm{x} = (v_{\rm max}, j)$, and the total likelihood is the product of likelihoods computed over all satellites as follows:
\begin{align} 
\begin{aligned}\label{eq:mom}
P&(\bm{D}|\, \bm{\Theta}) = P(\bm{D} |\, \bm{X}) =  \prod_{s=1}^{N_{\rm sat}} P({\bm d_s|\, \bm x_s}) \\ 
& = \prod_{s=1}^{N_{\rm sat}} N(j_s^{\rm obs} |\, j_s, \sigma_{j,s}^2) \times N(v_{{\rm max},s}^{\rm obs} |\, v_{{\rm max},s}, \sigma_{v_{{\rm max},s}}^2), \end{aligned}
\end{align}  
where $N_{\rm sat}$ is the total number of satellites and for each host halo in the sample, the subhalo with the highest value of $v_{max}$ is used in the $s=1$ term (i.e., the M33 analog) and so on, according to the rank order methodology discussed in Section \ref{subsec:prior}. We invoked the assumption that the measurements $\bm{d}$ of an individual satellite, conditional on their latent values $\bm{x}$, have no additional dependence on the halo mass $m$, such that $P(\bm d|\,\bm{x}, m) = P(\bm d|\,\bm x)$. This means that the measurement errors in $\bm{d}$ are independent of $m$. Following Bayes' theorem,
\begin{align}\label{eq:bayes1} 
   P(\bm{\Theta} |\,\bm{D}) \propto P(\bm{D}|\, \bm{\Theta}) P(\bm{\Theta}) ,
\end{align}
the posterior distribution for the mass of M31 is then computed using the likelihood, $P(\bm{D}|\, \bm{ \Theta})$, and the prior, $P(\bm{\Theta})$, via importance sampling as in \citetalias{patel17b}, \citetalias{patel18a}, and as described below.

Each observed satellite property is treated independently, as in previous work \citep[see also][]{busha11}. We note that this is not an ideal assumption as all satellite properties are derived as relative quantities with respect to M31. Furthermore, the nature of NGC~147 and NGC~185's relation is still uncertain. While our recent work \citep{sohn20} shows that these galaxies are not a binary pair orbiting M31 together, their PMs and LOS velocities are still very similar, implying some correlation between these puzzling dwarf ellipticals. A more detailed analysis including any correlations between satellite properties is beyond the scope of this and previous work. 

\subsection{Importance Sampling}
Generalizing from Section 3.2.1 of \citetalias{patel18a} to multiple satellites, Bayes's theorem is 
\begin{equation}\label{eq:bayes} 
 P(\bm{X}, m|\,\bm{D}) \propto P(\bm{D}| \, \bm{X}) \times P(\bm{X}, m|\, \bm{C}),
\end{equation}
where $\bm{C}$ denotes the dependence of the prior on the selection criteria ($\bm{C}$), the left-hand side is the posterior  distribution, $P(\bm{X}, m|\, \bm{C})$ is the prior probability distribution, and $P(\bm{D} | \, \bm{X})$ is the likelihood (see Eq. \ref{eq:mom}).

The posterior probability density function (PDF) is computed by drawing a set of samples of size $n$, from an importance sampling function. The importance sampling function is chosen to be the prior PDF, so importance weights are proportional to the likelihood. With these weights, integrals summarizing the target parameter $m$ are calculated as follows. The posterior expectation of a function $f(\bm{\Theta})$ of the latent properties is:
\begin{equation}
    \mathbb{E}[f(\bm{\Theta}) |\,\bm{D} ] = \int f(\bm{\Theta})\, P(\bm{X}, m|\, \bm{D}; \bm{C}) \,  d\bm{\Theta}. 
\end{equation}
If $f(\bm{\Theta})$ only depends on $m$, then the posterior expectation for $m$ is written
\begin{equation}\label{eq:posterior_1}
   \mathbb{E}[f(m) |\, \bm{D}] = \int  f(m) \, P(m|\, \bm{D}; \bm{C}) \, dm. 
\end{equation}
Using $n$ samples from the prior $P(m, \bm{X} |\,\bm{C})$, indexed as $j = 1,\ldots,n$, Equation \ref{eq:posterior_1} can be approximated as a Monte Carlo sum:
\begin{equation}\label{eq:posterior_2}
    \mathbb{E}[f(m) | \bm{D} ] \approx
    \sum_{j=1}^n  f(m^j) \, w_j,
\end{equation}
where $w_j$ are importance weights. When using only one satellite galaxy ($N_{\rm sat} = 1$), weights were calculated by 
\begin{align}
    w_j = \frac{P(\bm{d} |\, \bm{x}_j)}{\sum^n_j P(\bm{d} |\, \bm{x}_j)}, 
\end{align}
where $n$ is the total number of samples from the prior. To generalize this to multiple satellites ($N_{\rm sat} > 1$), the importance weights are now calculated as
 \begin{align}
    \begin{aligned}
    w_j = \frac{ \prod_{s=1}^{N_{\rm sat}}  P(\bm{d}_s| \, \bm{x}_{j,s})}{\sum^n_{i=1} \prod_{s=1}^{N_{\rm sat}}P( \bm{d}_s|\, \bm{x}_{i,s})}.
    \end{aligned}
    \label{eq:weights_nsat}
\end{align}
The derivation of these weights is given in Appendix \ref{app:derivation}.

Setting $f(\bm{\Theta})=m$ as in Eq. \ref{eq:posterior_2} gives the posterior mean value of M31's virial halo mass. The weights in Eq. \ref{eq:weights_nsat} are then used in a weighted kernel density estimate to compute posterior probability densities over $m$. See \citetalias{patel17b} and \citetalias{patel18a} for additional details on importance sampling and kernel density estimation.

For convenience, results are reported on a physical scale throughout as \Mvir$=X^{+U}_{-L}$ \Msun, where log$_{10}$X is the posterior mean of log$_{10}$\Mvir and [log$_{10}$(X -- L), log$_{10}$(X + U)] is the 68\% credible interval in log$_{10}$\Mvir. 

 \begin{figure}
    \centering
    \includegraphics[scale=0.46, trim=15mm 0mm 5mm 0mm ]{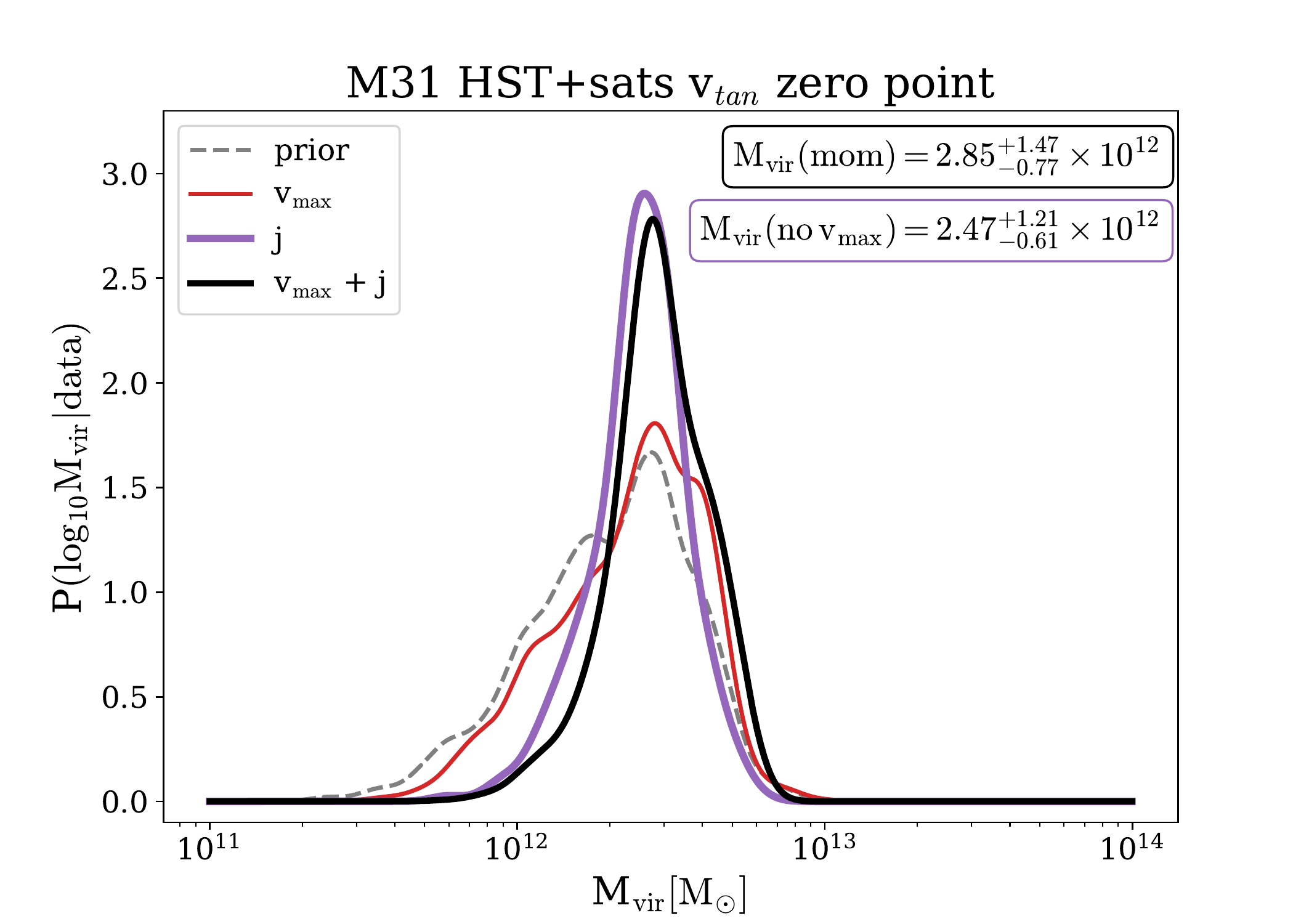}
    \caption{Posterior PDFs for the virial mass of M31 inferred using the properties of four M31 satellite galaxies. Observed satellite properties are relative to the M31 \textit{HST}+sats $v_{\rm tan}$ zero-point. Posterior PDFs are shown for the maximum circular velocity of the satellite's dark matter halo ($ v_{\rm max}$; red) and the total orbital angular momentum ($j$; purple). The dashed gray curve represents the underlying prior probability distribution given equal likelihood weights. The black curve corresponds to the total posterior PDF using $ v_{\rm max}+j$. The posterior mean M31 virial mass is $2.85^{+1.47}_{-0.77}\times10^{12}\, M_{\odot}$ (mom.). Uncertainties represent the 68\% credible intervals. The posterior mean M31 virial masses without the dependence on $v_{\rm max}$ is $2.47^{+1.21}_{-0.61}\times10^{12}\, M_{\odot}$ (mom.). Excluding $v_{\rm max}$ consistently yields a lower M31 \Mvir. }
    \label{fig:HSTresults}
\end{figure}

\section{Results: M31 Virial Mass Estimates}
\label{sec:result}
Using the observed data from Table \ref{tab:obsdata} and the statistical methods described above, we compute posterior PDFs for M31's virial mass. M31's mass is computed using the properties of four satellite galaxies simultaneously and is calculated with two different M31 tangential velocity zero-points.

\subsection{M31 \textit{HST}+sats
Tangential Velocity zero-point Results}
\label{subsec:result1}

Figure \ref{fig:HSTresults} shows posterior PDFs using the observed properties of M31 satellites for the \textit{HST}+sats $v_{\rm tan}$ zero-point as inputs to the likelihood functions (Eq. \ref{eq:mom}). The dashed gray line represents the underlying prior distribution assuming equal weights. The prior encompasses four orders of magnitude in virial mass from $10^{10}-10^{14}\, M_{\odot}$, with the highest probability regions spanning $2\times10^{11}-1\times10^{13}\, M_{\odot}$ (see Figure \ref{fig:prior}).

Individual posterior PDFs are shown for specific satellite properties, including satellite halo maximum circular velocity ($v_{\rm max}$; red) and satellite orbital angular momentum ($j$; purple). The posterior curves for the momentum method (black curve) indicate results using both satellite properties. The posterior mean M31 mass is \Mvir$=2.85^{+1.47}_{-0.77}\times10^{12}\, M_{\odot}$.

\subsection{M31 \textit{HST}+\textit{Gaia} DR2 Tangential Velocity zero-point Results}
\label{subsec:result2}
\begin{figure}
    \centering
    \includegraphics[scale=0.46, trim=15mm 0mm 5mm 0mm ]{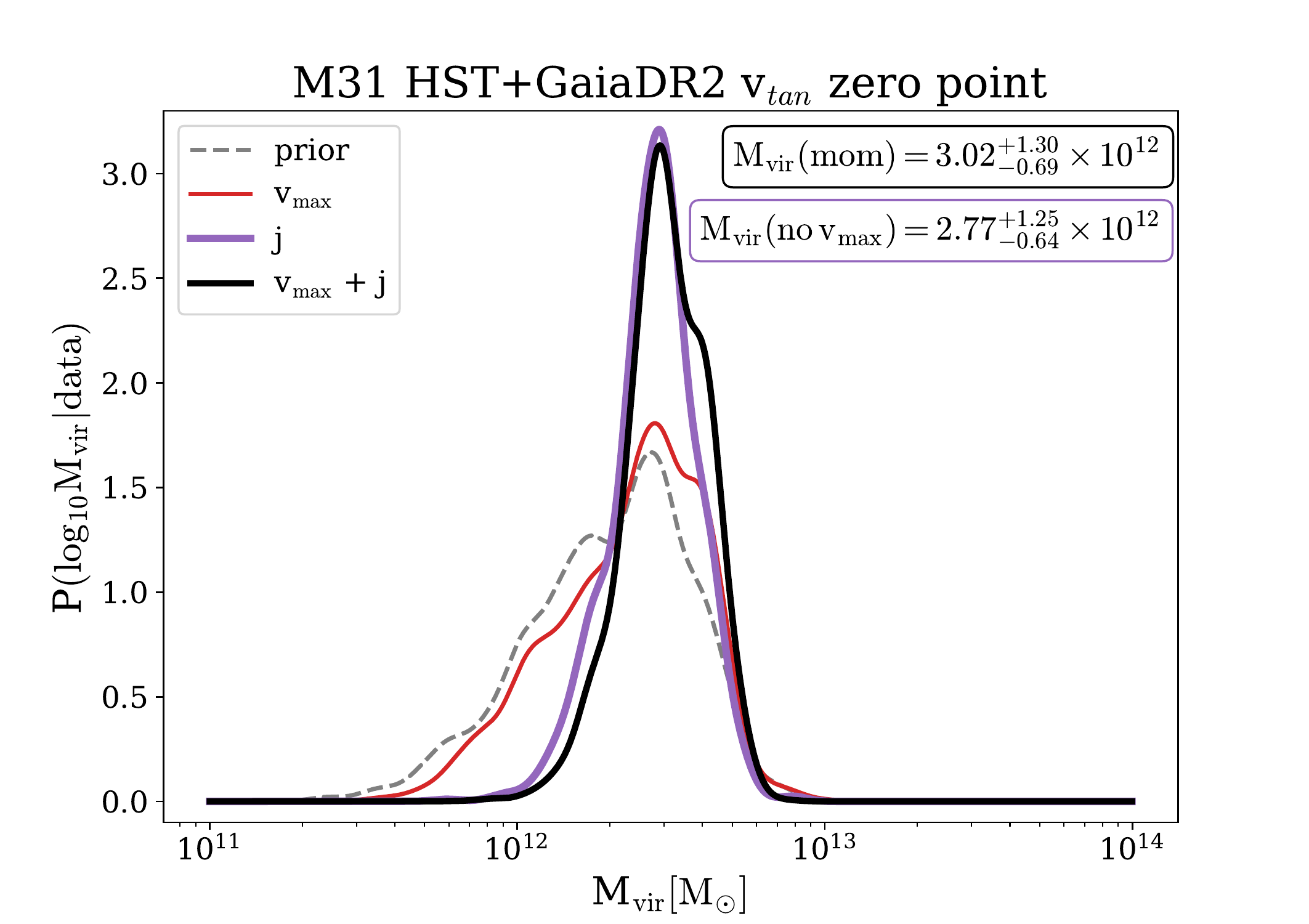}
    \caption{Same as Figure \ref{fig:HSTresults} using the satellite properties derived with the \textit{HST+Gaia} weighted average M31 $v_{\rm tan}$ zero-point. The posterior mean mass for M31 is \Mvir$=3.02^{+1.30}_{-0.69}\times10^{12}\, M_{\odot}$. When the likelihood dependence on $v_{max}$ is removed, the posterior mean M31 mass is \Mvir$=2.77^{+1.25}_{-0.64}\times 10^{12}\, M_{\odot}$.}
    \label{fig:Gaiaresults}
\end{figure}

Using satellite properties derived from the \textit{HST}+\textit{Gaia} M31 $v_{\rm tan}$ zero-point, we follow the same methodology to compute posterior PDFs for M31's virial mass. We find a posterior mean M31 \Mvir value of $3.02^{+1.30}_{-0.69}\times10^{12}\, M_{\odot}$, as illustrated in Figure \ref{fig:Gaiaresults}.

This $v_{\rm tan}$ zero-point results in significantly higher M31 \Mvir estimates as compared to the \textit{HST}+sats zero-point because M31's $v_{\rm tan}$ is $\approx 40$ km s$^{-1}$ higher when the \textit{HST}+\textit{Gaia} zero-point is adopted. The increase in M31's $v_{\rm tan}$ also propagates into the satellite's derived observed properties such that the total relative velocities and orbital angular momenta are also larger for all four satellites relative to the \textit{HST}+sats M31 tangential velocity zero-point (see Table \ref{tab:obsdata} and Appendix \ref{app:tng_results}). 

Comparing results in Figures \ref{fig:HSTresults} and \ref{fig:Gaiaresults} (see also Table \ref{tab:results}), it is clear that a more precise measurement of M31 and M33's PMs is crucial to further reduce the uncertainty in the mass of M31 with this statistical framework. Results from \textit{HST} GO-15658 (M31)/GO-16274 (M33; P.I. S.T. Sohn) will be key to such improvements as these data are expected to reach a precision three times smaller than what is currently possible with \textit{HST} or \textit{Gaia}.

\begin{table}[]
    \centering
    \begin{tabular}{|c|c|c|}\hline
    Satellites &  Momentum &  Momentum  \\ 
     & & (no v$_{\rm max}$) \\
    & \Mvir [10$^{12}\, M_{\odot}$] & \Mvir [10$^{12}\, M_{\odot}$]  \\ \hline
    \multicolumn{3}{|c|}{M31 HST+sats zero-point}  \\ \hline
    4 sats & $\bm{2.85^{+1.47}_{-0.77}}$ &  $2.47^{+1.21}_{-0.61}$ \\ \hline
    \multicolumn{3}{|c|}{M31 HST+Gaia DR2 zero-point} \\ \hline
     4 sats  &  $\bm{ 3.02^{+1.30}_{-0.69}}$ & $2.77^{+1.25}_{-0.64}$ \\ \hline
    \end{tabular}
    \caption{Estimated M31 virial masses using all four satellites. Posterior means are reported both with and without including $v_{\rm max}$ in the likelihood analysis. Including $v_{\rm max}$ generally results in masses that are 10-15\% higher than without $v_{\rm max}$. Our preferred masses are given in bold text. Uncertainties on the total mass of M31 across methods using all four satellites vary from 23-50\%, a significant improvement on the scatter in recent M31 literature masses, which ranges from approximately $0.7-3 \times 10^{12}\,M_{\odot}$.}
    \label{tab:results}
\end{table}

\subsection{Dwarf Ellipticals in a Cosmological Context}

\label{subsec:vmax}
The Too Big To Fail (TBTF) challenge describes the discrepancy between the circular velocity profiles of the most massive subhalos in dark matter-only simulations of LG-like environments (i.e., those subhalos expected to host luminous satellites) and the observed properties of LG dwarf satellites \citep{bk11b}. Though this was first observed for only MW satellites, other studies have shown that this phenomenon is also seen around M31 and even for dwarfs in the field \citep{tollerud14, gk14b}. 

\citet{garrisonkimmel19} is one of several studies \citep[e.g.,][]{fattahi16,brooks14,buck19} that have shown how including the evolution of both dark matter and baryons in a high-resolution simulation of LG-like environments can nearly eliminate the TBTF and missing satellites challenges. In addition to accounting for more realistic feedback processes, the two primary factors that relieve these tensions are enhanced tidal disruption and baryonic mass loss, neither of which can be accurately modeled if baryons are not included. 

There are still a few key outliers for which the circular velocity profiles of subhalos from simulations and observed dwarfs are not in agreement. However, for these outliers, the observed circular velocities are higher than their simulated counterparts, in other words, the opposite of the TBTF challenge is present. These outliers include the dwarf ellipticals NGC~147 and NGC~185, as well as the starburst galaxy IC~10 \citep{gk17a,garrisonkimmel19}. These high-density galaxies fail to exist across different simulation suites, including FIRE, NIHAO, and APOSTLE.
Here, we consider the consequences of using $v^{\rm obs}_{\rm max}$ in the estimation of M31's mass in light of this tension.

In this simple test, we calculate the mass of M31 using just the angular momentum of all four satellites. The purple curves in Figures \ref{fig:HSTresults} and \ref{fig:Gaiaresults} show the resulting M31 mass estimates when the term describing the normal distribution centered on $v^{\rm obs}_{\rm max}$ is eliminated from the joint likelihood function in Eq. \ref{eq:mom}. In both cases, the posterior mean M31 masses excluding the dependence on $v^{\rm obs}_{\rm max}$ are lower (see Table \ref{tab:results}). The credible intervals similarly decrease but still overlap with the results quoted in Sections \ref{subsec:result1} and \ref{subsec:result2} (see Table \ref{tab:results}). This suggests that the most likely mass of M31 given the presence of NGC~147, NGC~185, and IC~10 is $\sim 3\times 10^{12} \, M_{\odot}$, or in other words, it is difficult to reconcile the presence of galaxies with properties similar to NGC~147, NGC~185, and IC~10 around host halos $\lesssim 3 \times 10^{12} \, M_{\odot}$.

These results are supported by the fact that the FIRE simulations examined in \citet{garrisonkimmel19} only span the range of $0.8-1.54 \times 10^{12} \, M_{\odot}$ in virial mass. Based on our results, we would not expect high-density dwarfs to exist around host galaxies with such small halo masses -- i.e., host galaxies with halo masses below $\sim 3 \times 10^{12}\, M_{\odot}$ may not provide the appropriate conditions under which these galaxies form or evolve into the dwarfs we see today. Additionally, there may also be other factors at play that lead to the circular velocity discrepancy between observations and simulations for these dwarf outliers.
 First, the inferred maximum circular velocities of these galaxies may be overestimated. The rotation curves of NGC~147 and NGC~185 in particular have only measured out to 2-3 kpc \citep{geha10}, thus an approximation such as that described in Section \ref{subsubsec:ngcs} is necessary to construct circular velocity profiles out to larger radii. It could also imply that there is a missing piece in our understanding of galaxy formation at these dwarf mass scales. We encourage readers to consult \citet{garrisonkimmel19} for additional details.

\subsection{Limitations of the Method}
One limitation of the methodology discussed in \citetalias{patel17b} and \citetalias{patel18a} is a low effective sample size (ESS). Low ESS values occur when there are few samples in the prior with high importance sampling weights and thus few halo systems that dominate the posterior PDF. This could add additional sampling noise to the results discussed in Section \ref{sec:result}. Furthermore, this is one reason why we do not include results from the instantaneous method in this work, as we did in \citetalias{patel17b} and \citetalias{patel18a}. The instantaneous method often produces ESS values below 10, and thus the resulting credible interval is dominated by the dispersion across host halo properties for those 10 systems since random samples chosen from high-dimensional spaces (i.e., the instantaneous method needs to match three parameters for each of four satellites in a 12-dimensional space) tend to be farther apart than closer together.

To determine the magnitude of sampling noise using multiple satellites, we perform a bootstrap analysis by drawing, with replacement, 25 mock catalogs of the same size from the prior sample described in Section \ref{subsec:prior}. These mock catalogs are then used to estimate the mass of M31\footnote{Here we use only the \textit{HST}+sats M31 $v_{\rm tan}$ zero-point but similar results are expected for any $v_{\rm tan}$ zero-point. We also use the total likelihood function including $v_{\rm max}^{\rm obs}$ as given by Eq. \ref{eq:mom}.} following the methods in Section \ref{sec:methods}. The standard deviation of the posterior mean masses for 25 bootstrapped catalogs is $0.08 \times 10^{12} \, M_{\odot}$, confirming that the momentum method is robust against low ESS and effectively captures the true dispersion in posterior mean mass.

As a statistical approximation was implemented to combine individual satellite galaxy posteriors to estimate the mass of the MW in \citetalias{patel18a}, we tested the validity of the approximation by using the Bayesian framework to estimate the mass of 100 random halo systems using eight subhalos from the prior sample. Previously, we found that for 90\% of the randomly selected systems, the true host halo virial mass was recovered within two posterior standard deviations of the posterior mean (in dex), implying that the statistical approximation was under-performing. 

Since we remove the additional statistical approximation implemented in \citetalias{patel18a} to simultaneously constrain the host halo mass with four satellites, we expect to recover the true virial host halo mass for 95\% of randomly selected systems in the prior using this method if it is robust. For these mock tests, we randomly selected 100 halo systems and assign measurement errors on subhalo properties that are equivalent to the median of the observed data listed in Table \ref{tab:obsdata} (i.e., 10 km/s on $v_{\rm max}^{\rm obs}$, 20\% uncertainties on $r^{\rm obs}$ and $v_{\rm tot}^{\rm obs}$, 30\% uncertainty on $j^{\rm obs}$). Applying our methods, we find that the host virial mass is recovered within two posterior standard deviations of the mean for 96/100 randomly selected systems, implying that our M31 virial mass estimates in Section \ref{sec:result} are robust (i.e., our credible regions appropriately capturing the true mass). We note that in this process, we have excluded any likelihood weights corresponding to the random halo and any of its progenitors to ensure that the most highly weighted halos are not the tested halo itself (or the tested halo in previous snapshots of IllustrisTNG-Dark since repeated systems are allowed within our prior.)

Finally, one must consider the irreducible uncertainty associated with cosmic variance, or the imperfect correlation between subhalo properties and host halo mass. 
In \citetalias{patel17b} and \citetalias{patel18a} we quantified how well the uncertainty due to cosmic variance is captured by the credible intervals on the posterior mean M31 masses, finding that our methods accurately encompassed and even overestimated the uncertainty due to cosmic variance. Here, we repeat a similar exercise where a set of 25 halo systems are randomly selected from the prior and assigned measurement errors equivalent to those reported for the observed data corresponding to the satellites in our sample (see prior paragraph). For this exercise, we specifically choose systems where all four subhalos are located within 300 kpc of their host halo so the assigned measurement uncertainties accurately reflect the observed data.

We then calculate the host halo mass for all 25 systems and the root mean square (rms) error of the posterior log halo mass compared to the true log halo mass. This is compared to the average of posterior standard deviations (avg. $\sigma_{\rm post}$). The ratio of the rms error to the avg. $\sigma_{\rm post}$ deviations is $\rm \frac{rms}{avg. \,\sigma_{post}}$=0.62, confirming that cosmic variance is indeed accurately captured within our analysis framework. However, this value is smaller than that reported in \citetalias{patel17b} where we found $\rm \frac{rms}{avg. \,\sigma_{post}}$=0.87. The decrease from 0.87 to 0.62 is likely due to a smaller ESS when matching the properties of four satellites as opposed to just one satellite as in \citetalias{patel17b}. The value $\rm \frac{rms}{avg. \,\sigma_{post}}$=0.62 can be interpreted such that our uncertainties are ``overconfident" in accounting for cosmic variance by $\sim$38\% compared to the actual rms errors.

\section{Discussion}
\label{sec:discussion}

\subsection{Comparison to Literature and Previous Work}
\label{subsec:lit}
\begin{figure*}
    \centering
    \includegraphics[scale=0.55, trim=0mm 6mm 0mm 5mm]{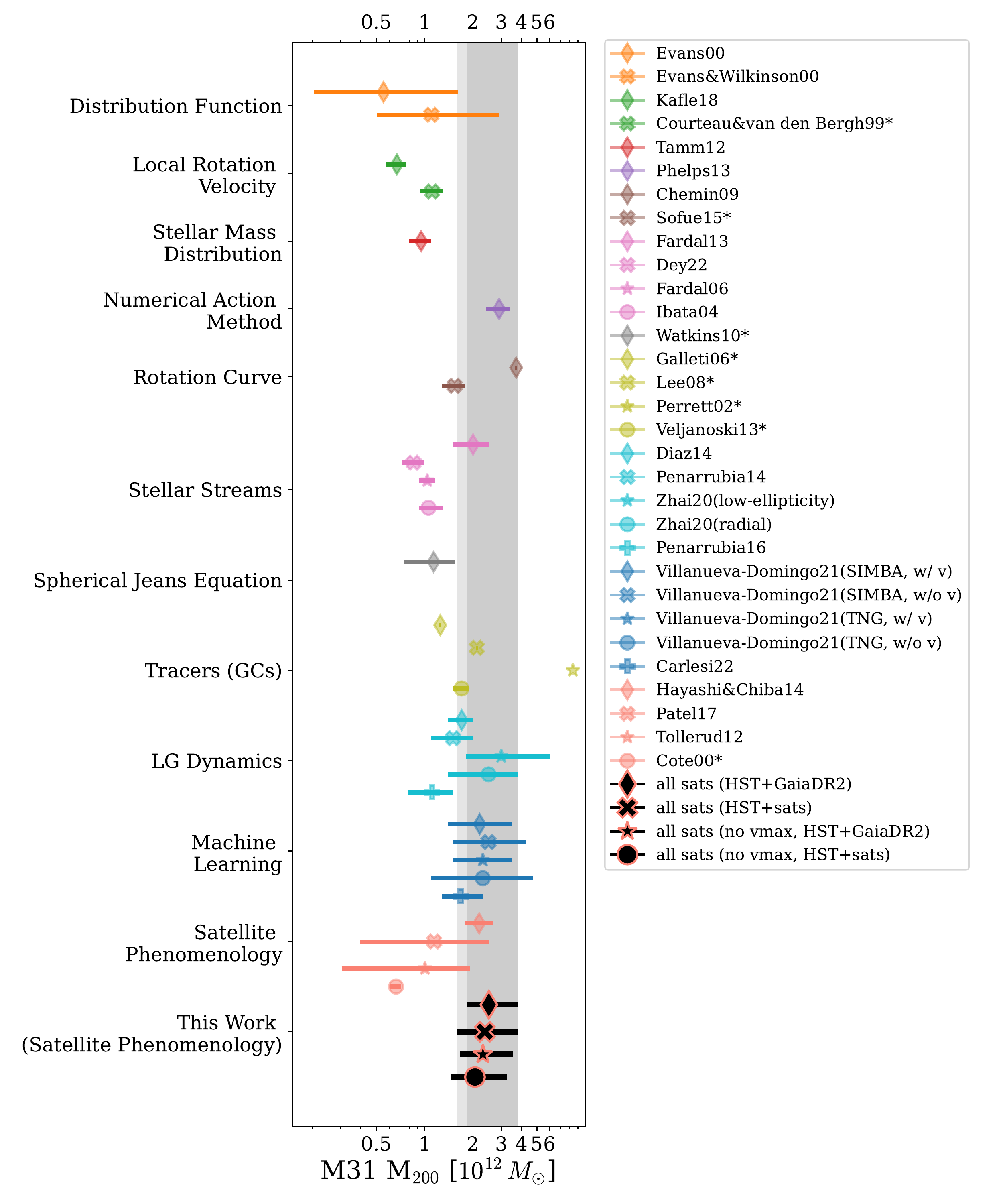}
     \caption{A selection of literature M31 mass estimates from the last two decades. The method by which masses were determined is marked on the left and points are grouped by symbols with the same color. Literature masses have been converted to $M_{200}$ assuming NFW profiles and errors typically correspond to 68\% confidence intervals. If only statistical errors are reported, labels are marked with an asterisk. Results from this work are given by black symbols outlined in salmon. The two mass estimates using all four satellites including $v_{\rm max}$ are shaded in grey. The extents of the gray regions span $M_{200} = [1.60, 3.81] \times 10^{12}\, M_{\odot}$ or \Mvir$= [2.08, 4.32] \times 10^{12} \, M_{\odot}$. }
    \label{fig:litmasses}
\end{figure*}

Nearly a dozen different techniques have been used to estimate the total mass of M31, the MW, and the combined mass of the LG over the last two decades. In Figure \ref{fig:litmasses}, we illustrate a selection of literature masses \citep{evans00, evans_wilkinson00, kafle18, courteau_vandenbergh99, tamm12,phelps13,chemin09,sofue15,fardal13,dey22,fardal06,ibata04,watkins10,galleti06,lee08,perrett02,veljanoski13,diaz14,penarrubia14,zhai20, penarrubia16, villanueva_domingo21, carlesi22, hayashi14, patel17b, tollerud12, cote00} to provide context for the results of this work. Literature data points are from the following studies: 

For Figure \ref{fig:litmasses}, literature masses have been converted to $M_{200}$ assuming NFW profiles and the mass-concentration relation for field halos from \citet{klypin11}, similar to Figure 1 of \citet{wang20}. Here, $M_{200}$ refers to the mass enclosed within $R_{200}$, the radius inside which the density is 200 times the critical density of the Universe. Uncertainties correspond to the 68\% errors, which are taken directly from the literature for a majority of cases. In some situations \citep[e.g.,][]{carlesi22}, original errors, such as the first and third quartiles, are adopted. For \citet{phelps13}, 95\% errors are converted to 68\% errors assuming the errors are Gaussian in linear space. We note with an asterisk the literature masses where only statistical errors are reported. We caution that the total error is underestimated (i.e., observational and systematic uncertainties are not included) for these literature masses and therefore should not be directly compared to the precision of other mass estimates, such as those from this work.

In Figure \ref{fig:litmasses}, our results fall into the ``satellite phenomenology" category and are marked as black symbols outlined in salmon (see Table \ref{tab:results}). The gray shaded regions encompass the two results we find using all four satellites and each of the M31 $v_{\rm tan}$ zero points. Our masses encompass the region spanning $M_{200} = [1.60, 3.81] \times 10^{12} \, M_{\odot}$ or \Mvir$= [2.08, 4.32] \times 10^{12} \, M_{\odot}$. 

Our posterior mean masses are most consistent with the results of \citet{villanueva_domingo21, zhai20, hayashi_chiba14, fardal13, phelps13} who have used machine learning (blue symbols), LG dynamics (cyan symbols), satellite phenomenology (salmon symbols), stellar streams (pink symbols), and the numerical action method (purple symbol). Of these, the results from \citet{villanueva_domingo21} and \citet{phelps13} also reach similar precision to our results.

In particular, it is especially interesting to compare the precision and values of our new results with those from \citetalias{patel17b}. Previously by considering only M33, we found \Mvir$=1.37^{+1.39}_{-0.75}\times10^{12}\, M_{\odot}$ using the \textit{HST}+sats M31 $v_{\rm tan}$ zero-point and the Illustris-Dark simulation, whereas upon expanding the sample size to include three additional satellites and upgrading to IllustrisTNG-Dark, we find \Mvir$=2.85^{+1.47}_{-0.77}\times10^{12}\, M_{\odot}$ (\textit{HST}+sats). Though the posterior mean masses differ as a function of satellite sample size, they are consistent with one another within the credible intervals. Comparing the precision of these values, using only M33 yields an uncertainty of $\sim$50-120\%, while four satellites yield only $\sim$23-50\% (regardless of which M31 $v_{\rm tan}$ zero-point is adopted). We conclude that quadrupling the satellite sample yields a reduction of more than $\sim$50\% in the uncertainty on the mass of M31\footnote{The systematic offset between results from Illustris-Dark and IllustrisTNG-Dark is only $\sim$5\% (see Appendix \ref{app:tng_results} for details).}.

\subsection{Local Group Mass}
\label{subsec:lg}
The mass of M31 is intimately tied to the dynamical history of the LG and our understanding of the LG in a cosmological context. Often, the mass of the MW, M31, or both are used to identify analogs in cosmological simulations with which the assembly of the LG is studied. Orbital modeling has also been used to understand the accretion history and trajectory of substructure in the LG, however, the assumed masses of the MW and M31 are one of the primary causes for large uncertainties \citep[e.g.,][]{patel17a, li21, battaglia22}. Understanding what fraction of the LG's mass is attributed to M31 is crucial to such studies and therefore, we briefly discuss how the results presented in this analysis compare with recent LG mass estimates.

One of the most notable methods used to estimate the combined mass of the LG is the Timing Argument (TA) based on the fact that the MW and M31 are currently approaching one another, having overcome the effects of cosmic expansion during the early Universe due to their strong gravitational attraction. While the TA has been revised over time as more precise data have become available for LG galaxies, literature masses for the LG have varied between $2.5-5 \times 10^{12} \, M_{\odot}$ through the 2010s.  

Since then several studies have taken the TA one step further by including the influence of the LMC given its substantial mass \citep[at least $\sim$ 10\% the mass of the MW;][]{penarrubia16, benisty22, chamberlain22}. The LMC is expected to displace the MW's disk and inner halo over time, and impart a reflex motion on the MW \citep[e.g.][]{gomez15, garavitoc19, petersen20, cunningham20, gc21}. Furthermore, \cite{petersen21} recently measured the motion of the inner MW halo relative to the outer halo, or ``travel velocity", confirming that the gravitational influence of the LMC is critical to the dynamics of the LG. As such, these effects are important to consider in the context of the TA, which has traditionally only included the MW and M31. 

\cite{penarrubia16} included the effect of the LMC by assuming that the MW and LMC could be treated as a two-point system and that M31 is in orbit around the MW-LMC barycenter. By modeling the three galaxies and other Local Volume galaxies as a dynamic system, they simultaneously fit for distances and velocities of all galaxies while leaving the LMC's mass as a free parameter in a Bayesian fashion. In doing so, individual masses are derived for each of the MW, M31 (see Fig. \ref{fig:litmasses}), the LMC, and the total LG mass for which they find $2.64^{+0.42}_{-0.38} \times 10^{12} \, M_{\odot}$.  

\cite{benisty22} revisited the TA allowing for a cosmological constant and including the influence of the LMC. All galaxies are considered as extended mass distributions rather than as point masses as in \citet{penarrubia16}. The TA including the LMC yields a mass of $5.6^{+1.6}_{-1.2} \times 10^{12} \, M_{\odot}$, approximately 10\% lower than the resulting LG mass when only the pure TA (i.e, no LMC) is computed. Accounting for cosmic bias and scatter in addition to the influence of the LMC reduces the LG mass to $3.4^{+1.4}_{-1.1} \times 10^{12} \, M_{\odot}$.

\cite{chamberlain22} reevaluate the TA including the measured travel velocity in the equations of motion describing the two-body MW-M31 system. They use three different sets of PM and distance data for M31, including data that overlaps with the observed data used in this work. They find LG masses of $3.98^{+0.63}_{-0.47} \times 10^{12} \, M_{\odot}$ using an M31 distance from \citet{vdmG08} and the \textit{HST}+sats M31 PM; $4.05^{+0.51}_{-0.34} \times 10^{12} \, M_{\odot}$ using a distance from \cite{li21} and the \textit{HST}+sats M31 PM; and finally, an LG mass of $4.54^{+0.77}_{-0.56} \times 10^{12} \, M_{\odot}$ using the distance from \cite{li21} and \textit{Gaia} eDR3 PM from \cite{salomon21}. \citet{chamberlain22} do not account for a cosmological constant and cosmic bias like \citet{benisty22}, yet both results consistently find that including the influence of the LMC yields an LG mass that is approximately 10-20\% lower than the pure TA with only the MW and M31 with respect to each of their methods. 

All three TA studies including the influence of the LMC neglect the influence of M33, the third most massive galaxy in the LG, and the most massive satellite of M31. This is because the M31 reflex motion owing to the passage of M33 is unknown and subsequently a travel velocity has yet to be measured. If M33 is truly on first infall as predicted by \citet{patel17a}, the M31 reflex motion is expected to be small relative to the MW reflex motion, however, these predictions are based on M31 masses smaller ($1.5-2 \times 10^{12} \, M_{\odot}$) than those reported in this work. Therefore, it will be necessary to revisit the orbital history of M33 with a massive M31 to predict an accurate M31 reflex motion (Patel et al., in prep.). Preliminary results following the methodology of \citet{patel17a} indicate that for a $3 \times 10^{12} \, M_{\odot}$ M31, M33 passes through pericenter $\sim$4 Gyr ago at a distance of $\sim$100-200 kpc. Once a prediction for the M31 reflex motion exists, this value or a measurement of the M31 disk travel velocity can be incorporated into the TA as in \cite{chamberlain22}.

Our M31 masses are in agreement with the LG masses found in \citet{chamberlain22} if we assume a MW mass of $0.96 \times 10^{12} \, M_{\odot}$ \citepalias{patel18a}. Adding our M31 masses to previous MW mass results, we find a total LG mass of $\sim 4\times10^{12}\,M_{\odot}$. Outside of the TA method, other studies have used cosmological simulations, machine learning, and LG dynamics to constrain the total mass of the LG. A selection of these masses has been compiled in Table 4 of \cite{chamberlain22} and Fig. 4 of \cite{benisty22}. Assuming our \citetalias{patel18a} mass for the MW results from this work are most consistent with \cite{mcleod17, zhai20, lemos21} who find an LG mass of at least $\gtrsim 4\times 10^{12}\, M_{\odot}.$

\subsection{M31 Stellar Mass Fraction}
\label{subsec:smhm}
\begin{figure}
    \centering
    \includegraphics[scale=0.53, trim=10mm 5mm 15mm 0mm ]{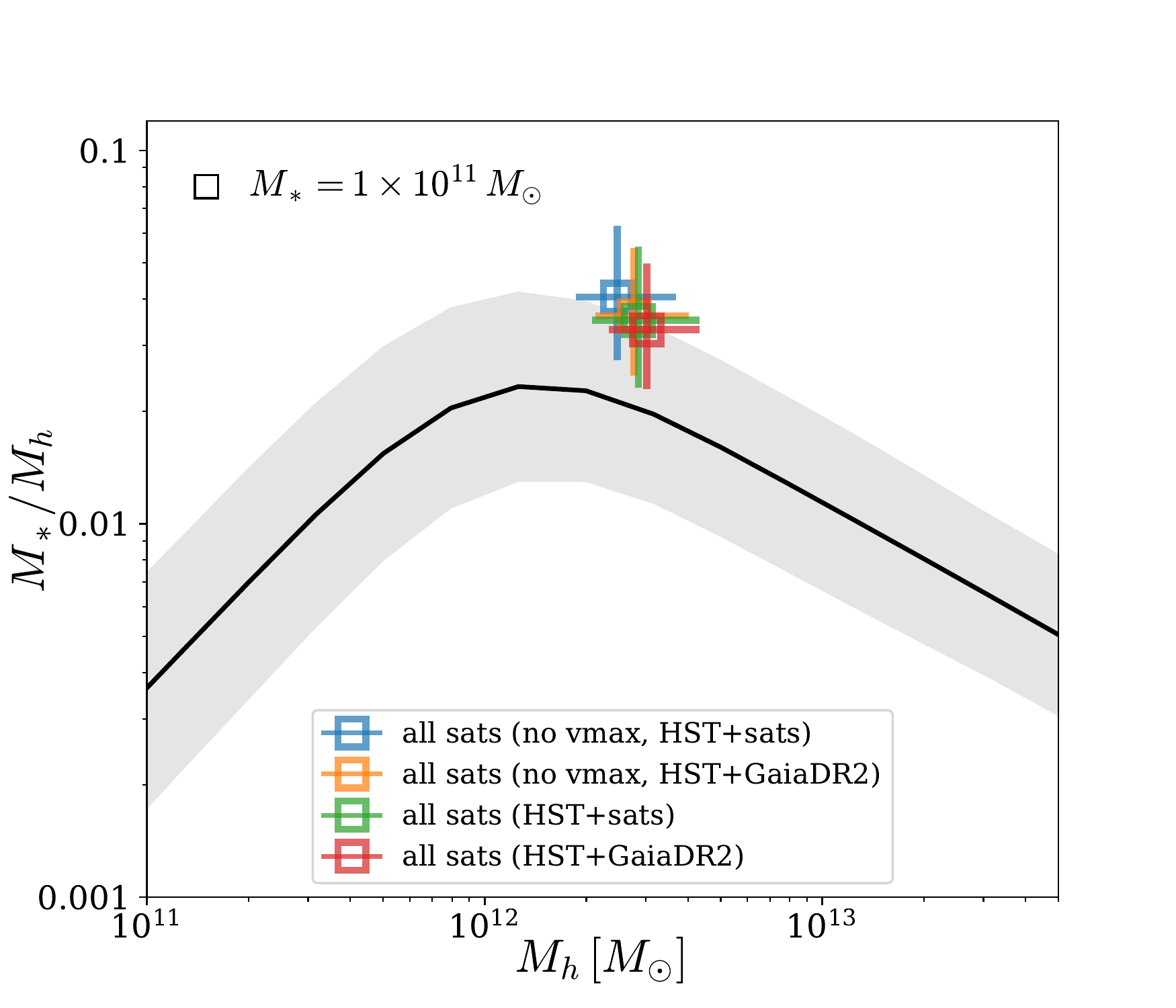}
    \caption{The median stellar mass--halo mass relation (SMHM) from the Universe Machine \citep{behroozi19} is illustrated in black. The gray shaded region encompasses the 1$\sigma$ variance around the median SMHM relation. Square data points with error bars reflect the observed positions of M31 adopting $M_*=10^{11}\,M_{\odot}$ and the results listed in Table \ref{tab:results}. The stellar mass fraction corresponding to the results presented here is consistent with the upper limits of the SMHM relation at fixed halo mass.}
    \label{fig:smhm}
\end{figure}

The stellar mass--halo mass relation (SMHM) relation, which describes the correlation between galaxy stellar mass and dark matter halo mass, is often used to place galaxies in a cosmological context \citep[e.g.,][]{moster13, behroozi13b, brook14, gk14a, gk17a, behroozi19}. Previously, M31 has been known to lie far outside the scatter in the SMHM relation \citep[e.g.,][]{mcgaugh21}, so here, we use the halo masses reported in this work to place M31 on the SMHM relation. 

Figure \ref{fig:smhm} shows the SMHM relation from the Universe Machine \citep{behroozi19} for a range of halo masses corresponding to the approximate extents of our prior sample. The gray shaded region encompasses the scatter around the median SMHM as well as the associated statistical uncertainties. The observed stellar mass--halo mass ratio is derived from our results assuming $M_*=10^{11}\,M_{\odot}$ \citep{tamm12, sick15}. Uncertainties on $M_*/M_h$ include an additional 0.1 dex uncertainty on $M_*$ to account for the systematic error in converting luminosity to stellar mass and to most accurately reflect the uncertainties for different methods of modeling M31's stellar mass. Our results are consistent with the upper end of the SMHM relation, suggesting that a high mass ($\geq 3 \times 10^{12} \, M_{\odot}$) is cosmologically most favorable (i.e., our \textit{HST+Gaia} result; red square). 

\subsection{Reconciling a High M31 Mass and Observed M31 Properties}
\label{subsec:rec}
Our statistical framework is intentionally designed to minimize the criteria that determine which host halos are used as draws from the prior sample. As discussed in Section \ref{subsec:prior}, draws from the prior sample are only restricted by the minimum virial mass of host halos at $z\approx0$, selected as all halos with a minimum mass of $10^{10}\, M_{\odot}$. The only other criterion applied to host halos themselves is that their peak circular velocity, or $v_{\rm max}$, is less than 250 \kms. 

We set the upper $v_{\rm max}$ limit to 250 \kms because this is the approximate peak of M31's observed rotation curve \citep{corbelli10}. In practice, this is a generous upper limit since IllustrisTNG-Dark only follows the evolution of dark matter, and therefore just the dark matter halo's contribution to the observed rotation curve is relevant. Based on model decompositions of M31's rotation curve \citep{patel17a}, we approximate that the peak velocity of M31's dark matter halo is $200\pm20$ \kms, where the uncertainty in part depends on whether the halo has been adiabatically contracted or not. Given this $v_{\rm max}$, we select only those host halos in the prior sample with $v_{\rm max}=180-220$ and find \Mvir= $2.52\pm0.70 \times 10^{12} \, M_{\odot}$, illustrating a fairly tight correlation between $v_{\rm max}$ and \Mvir. This is similar to the results reported in Section \ref{sec:result} confirming that even a high mass M31, as reported in this work, can be reconciled with observations of the M31 rotation curve.
  
Another approach to reconciling our M31 mass results with the observed properties of the M31 system is to compare the radial velocity dispersion of all known M31 satellites to the radial velocity dispersion calculated using the first 30 subhalos in each host halo representing a draw from the prior distribution.

Radial velocities for individual M31 satellites are taken primarily from \cite{collins13} and our Table \ref{tab:obsdata}. These values are corrected to the M31-centric frame by subtracting the M31 radial velocity. Taking the standard deviation of these M31-centric radial velocities gives $\sigma_{rad} \approx 122\pm3$ \kms, which implies \Mvir = $2.88\pm1.24 \times10^{12}\, \,M_{\odot}$. The standard deviation in \Mvir is substantial as the correlation between \Mvir and $\sigma_{rad}$ is broad at a given $\sigma_{rad}$. The virial mass implied by the radial velocity dispersion of $\sim$30 M31 satellites is therefore consistent with, but much broader than the results reported in Section \ref{sec:result}. We note that mass estimation techniques relying on Jeans modeling \citep[e.g.,][]{watkins10}, for example, often give much smaller mass uncertainties, which implies that non-equilibrium host halos and satellite anisotropy of the satellite systems play a key role in the IllustrisTNG-Dark simulations.

\subsection{Implications for the M31 System}
\label{subsec:implications}
Large uncertainties in the masses of the MW and M31 are one of the main causes of significant uncertainties in modeling the backward orbital trajectories of halo substructures \cite[see][for example]{dsouza22}. We have previously demonstrated how the orbits of the LMC and M33, the most massive satellites with respect to the MW and M31, change when the masses of their host galaxies are varied in \citet{patel17a}. In particular, we presented a modified orbital history for M33 where it is statistically expected to be on first infall. These conclusions are based on an M31 mass of $[1.5, 2] \times10^{12} \, M_{\odot}$ and distances, LOS velocities, and PMs similar to those derived from the M31 \textit{HST}+sats zero-point used in this work. 

Other studies have adopted positions and velocities derived from different sets of PMs and/or assumed various M31 (and M33) masses and modeling techniques. For example, \cite{watkins13} have tabulated a census of orbital properties for all known M31 satellites assuming M31 $M_{\rm vir} = 1.1-1.9 \times 10^{12} \, M_{\odot}$ even in the absence of full 6D phase space data for satellites. This provides a uniform reference point for future orbit comparisons using a higher range of M31 masses. 

Additionally, \cite{mcconnachie09} find an M33 orbital history where M33 has a close passage around M31 at 2-3 Gyr ago, however, their adopted M31 mass is $2.5\times 10^{12} \, M_{\odot}$ and their M33 mass is only 30\% of the \cite{patel17a} adopted value. Similarly, \cite{putman09} find an orbit in agreement with that of \cite{mcconnachie09} modeling M33 as a point mass and using a total mass of $2\times 10^{12} \, M_{\odot}$ for M31. Most recently, \cite{teppergarcia20} adopted gravitational potentials and masses nearly identical to those used in \cite{patel17a} and found an orbital history consistent with the \citet{patel17a} high mass M31 ($2 \times 10^{12}\,M_{\odot}$) results where M33 makes a 50-100 kpc pericentric passage around M31 at $\sim$ 6-6.5 Gyr ago. 
 
\citet{sohn20} presented first orbital solutions using HST PMs for NGC~147 and NGC~185, however, they assumed low M31 masses relative to those presented in this work \citep[they adopted the same values as in][]{patel17a}. These results disproved the suggestion that these two galaxies are a binary system and found evidence for the formation of NGC~147's tidal tails from a close encounter with M31. A higher M31 mass would make it even less likely that NGC~147 and NGC~185 are a binary pair since the tidal field owing to M31 would be even stronger. These orbits will be revisited in future work. 

It is clearly evident that the mass of M31 (and M33) is key to accurate interpretations of the accretion history of the entire M31 system, potentially even into the M33 satellites regime as has been shown for LMC satellites \citep[e.g.,][]{patel20, garavitoc19}. In future work, we will quantify the consequences of a high M31 mass on M31's satellite population. It will be especially interesting to see whether evidence of group infall increases or decreases assuming a higher M31 mass with respect to the results of \citet{watkins13}.

It is worth noting that a higher M31 mass would also affect interpretations of its merger history, the formation of prominent stellar structures, and the evolution of globular clusters \citep[see][for a census of M31 substructure]{mcconnachie18}. Of recent interest in particular is whether M31 recently underwent a major \citep[e.g.,][]{dsouza18, hammer18} or a minor \citep[e.g.,][]{ibata04, font06, fardal06} merger. While addressing these topics is beyond the scope of this work, it will be crucial to understand the impact of a high mass on the entire M31 system as more data becomes available from \textit{HST, Gaia, DESI, JWST,} and more.

\section{Summary and Conclusions}
\label{sec:summary}
Building on the Bayesian framework previously used to estimate the mass of the MW with multiple satellite galaxies \citepalias[][]{patel18a}, here we have used the IllustrisTNG-100-Dark simulation in combination with observed properties of four M31 satellite galaxies (M33, NGC~185, NGC~147, and IC~10) to constrain the total mass of M31. These four M31 satellites are the only M31 satellites with available PMs from \textit{HST} and/or \textit{Gaia}. Throughout this work, we present two sets of results for observed satellite properties derived from \textit{HST}-based M31 PMs \citep[denoted as \textit{HST}+sats;][]{vdm12ii} and \textit{HST}+\textit{Gaia}-based M31 PMs \citep[denoted as \textit{HST}+\textit{Gaia};][]{vdm19}.  

We emphasize the use of dynamical satellite properties, such as orbital angular momentum, to constrain the mass of host galaxies, as we have shown such techniques to be the most robust against varying orbital configurations. \citepalias[see][]{patel17b}. The main conclusions of this work are summarized below.

\begin{enumerate}[leftmargin=*]
    \item Using the modified Bayesian framework outlined in \S \ref{sec:methods} and the orbital angular momentum for four satellite galaxies, we find two preferred estimates for M31: \Mvir$= 2.85^{+1.47}_{-0.77}\times10^{12}\, M_{\odot}$ (using the M31 \textit{HST}+sats $v_{\rm tan}$ zero-point) and \Mvir$=3.02^{+1.30}_{-0.69}\times10^{12}\, M_{\odot}$ (\textit{HST}+\textit{Gaia} $v_{\rm tan}$ zero-point; see \S \ref{sec:result}). 
    
    \item Including $v_{\rm max}$ in the likelihood function results in masses that are 10-15\% higher than without $v_{\rm max}$. Our results are robust against the Too Big To Fail challenge, however, this suggests that M31 must be at least as massive as $\sim 3\times 10^{12} \, M_{\odot}$ to host satellites with properties similar to NGC~147, NGC~185, and IC~10, which are typically outliers in simulations following the evolution of both dark matter and baryons (see \S \ref{subsec:vmax}). 
    
    \item For both M31 $v_{\rm tan}$ zero-points uncertainties range from 23-50\% compared to 50-120\% when only one satellite (M33) is used \citepalias{patel17b}. By using a sample with four times more satellites (see \S \ref{subsec:lit}), the uncertainties on M31's mass are more than halved. When 6D phase space information is available for all 35 M31 satellites through \textit{HST} GO-15902 (PI: D. Weisz), \textit{HST} GO-16273 (PI: S.T. Sohn), and JWST GTO 1305 (PI. R. van der Marel); we expect to reach uncertainties below 20\% \citep[see also][]{li17}.
    
    \item Comparing to literature M31 masses, the precision and numerical values we find are closest to studies using LG dynamics, the numerical action method, and machine learning. The advantage of our method is that it does not require strong assumptions about the properties of M31's halo or about the orbital configuration of satellites. Of the M31 masses compiled from analyses that do account for both observed measurement errors and systematic uncertainties, our results are amongst those with the highest precision (see \S\ref{subsec:lit}).
    
    \item Our M31 mass results are consistent with recently revised estimates for the total mass of the LG ($4-4.5 \times 10^{12}\, M_{\odot}$), assuming the mass of the MW is $\approx 10^{12} \, M_{\odot}$ \citepalias[][see \S \ref{subsec:lg}]{patel18a}.
    
    \item Our observed M31 stellar mass--halo mass fractions are consistent with the upper limits of the median SMHM relation at fixed halo mass, indicating that a high halo mass ($\geq 4 \times 10^{12} \, M_{\odot}$) is cosmologically most favorable (see \S \ref{subsec:smhm}).

    \item Our M31 masses are consistent with the observed properties of M31, including the observed rotation curve and the radial velocity dispersion of nearly all M31 satellites, implying that a high mass M31 ($\sim 3\times 10^{12}\, M_{\odot}$) is plausible given velocity measurements for stars in the M31 disk and in tracer populations (see \S\ref{subsec:rec}). 
    
    \item The implications of an M31 mass $> 2.5 \times 10^{12} \, M_{\odot}$ are expected to be substantial, particularly in the context of orbital modeling for substructures throughout M31's halo. This will be the subject of future work (see \S \ref{subsec:implications}).
\end{enumerate}

To utilize the abundance of satellite phase space information that has been published for MW satellites since \citetalias{patel18a} and to prepare for more M31 satellite phase space information, it will be necessary to move beyond the combination of large-volume cosmological simulations and Bayesian statistics to effectively further constrain the precise masses of the MW and M31. 

We have discussed how low satellite statistics in N-body simulations result in higher relative uncertainties in the inferred host halo masses. Neural networks overcome the problem of low statistics by training on satellite properties for a wide range of halos instead of only those most similar to the MW or M31, yielding greater constraining power. These modern methods also accommodate correlated satellite properties (e.g., infalling subhalos will have satellites of their own) without biasing the inferred host halo masses. Neural networks also have the advantage of self-consistently including arbitrary additional information (e.g., larger-scale environment or gas rotation velocities) to improve halo mass recovery. This will be the topic of upcoming work (Hayati et al., in prep.) and the results are expected to simultaneously improve our understanding of the MW and M31's mass in addition to constraining other galaxy and halo properties.



\software{This work has been possible thanks to \texttt{astropy} \citep{astropy}, \texttt{numpy} \citep{numpy}, \texttt{scipy} \citep{scipy}, and \texttt{matplotlib} \citep{matplotlib}. The IllustrisTNG data are publicly available at \url{https://www.tng-project.org/data/}.}




\section*{Acknowledgements}
E.P. acknowledges support from HST GO-15902 and HST AR-16628. Support for GO-15902 and AR-16628 was provided by NASA through a grant from the Space Telescope Science Institute, which is operated by the Association of Universities for Research in Astronomy, Inc., under NASA contract NAS 5-26555. K.S.M. acknowledges funding from the European Research Council under the European Union’s Horizon 2020 research and innovation program (ERC Grant Agreement No. 101002652). E.P. thanks Tony Sohn for producing the proper motion Monte Carlo files for this analysis, Alessandro Savino for graciously sharing distance moduli in advance of publication. E.P. would also like to thank Mike Boylan-Kolchin and Peter Behroozi for informative discussions which have helped improve the quality and context of this work. Additionally, Gurtina Besla, Roeland van der Marel, Dan Weisz, Nico Garavito-Camargo, Laura Watkins, and Katie Chamberlain have provided generous feedback on this manuscript.

\bibliography{postdoc_refs}{}
\bibliographystyle{aasjournal}


\appendix
\section{Estimating M31's Mass with M33 Using IllustrisTNG-Dark vs. Illustris-Dark}
\label{app:tng_results}

\begin{figure*}
\begin{center}
    \includegraphics[scale=0.46, trim=15mm 0mm 5mm 0mm ]{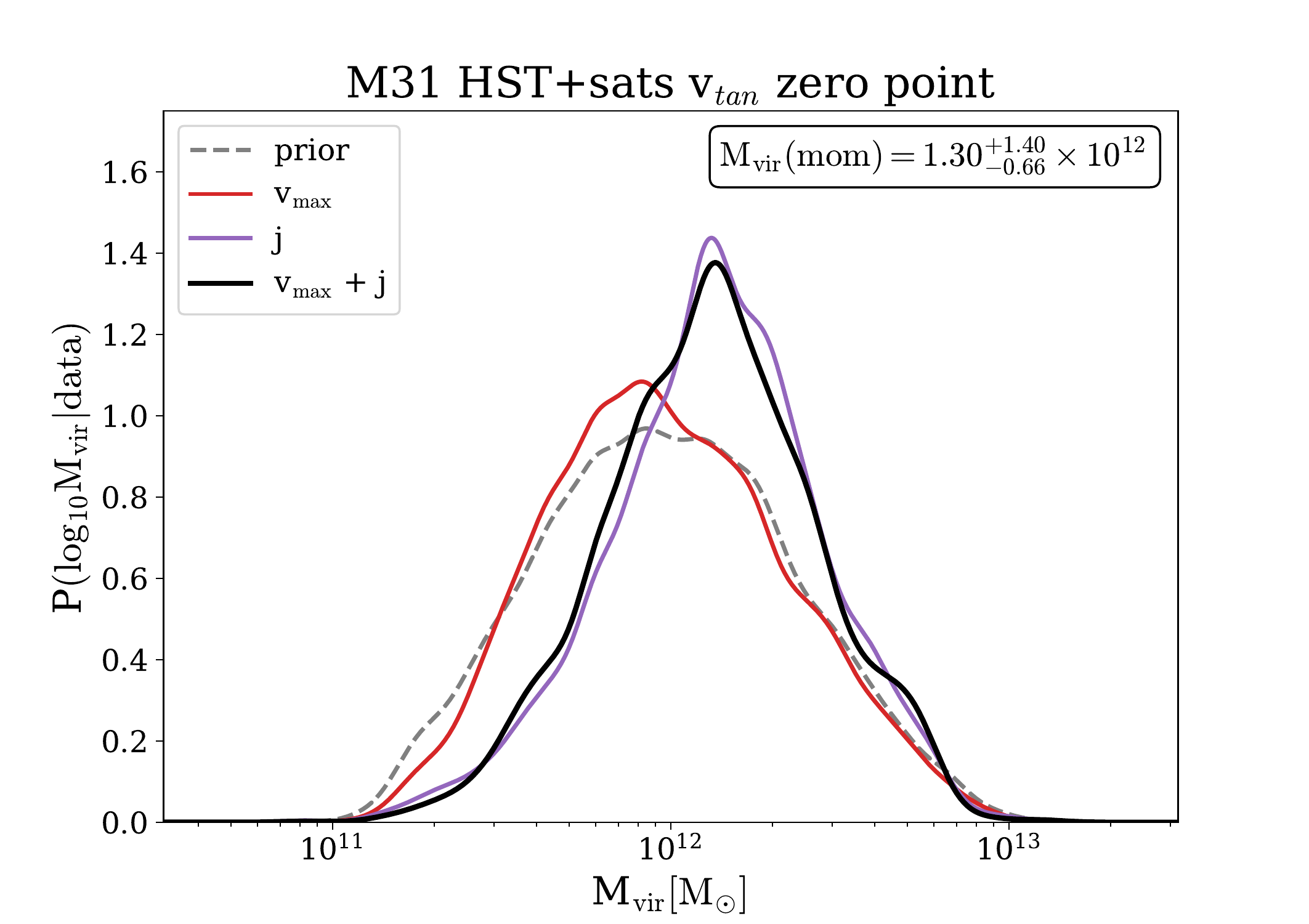}
        \includegraphics[scale=0.46, trim=15mm 0mm 10mm 4mm]{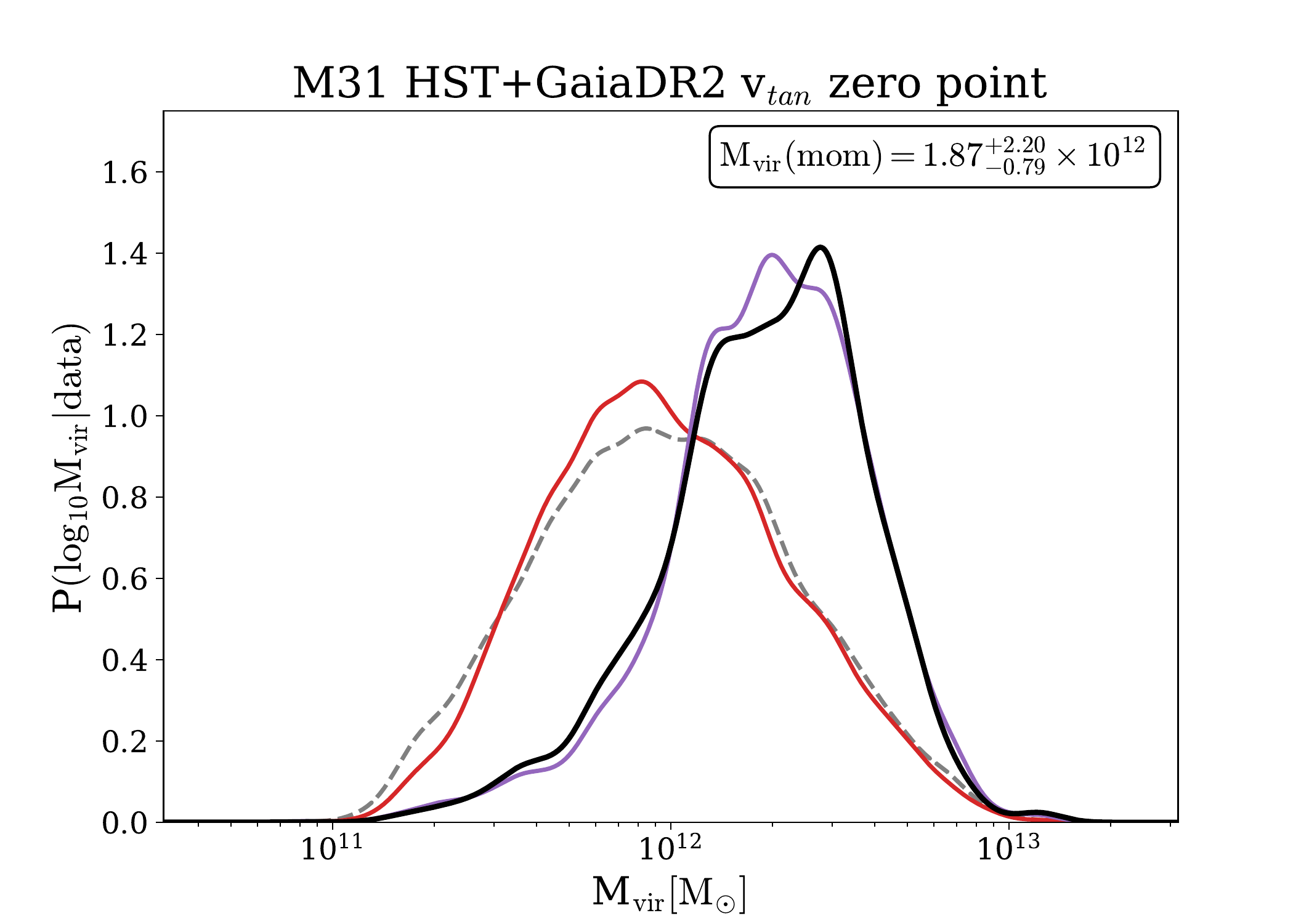}
    \caption{Posterior distributions for the virial mass of M31 using the properties of M33 (see Table \ref{tab:obsdata}) and the IllustrisTNG-Dark simulation. \textbf{Left:} The mass resulting from the momentum (black) method using the \textit{HST}+sats M31 $v_{\rm tan}$ zero-point. Posterior PDFs are also shown for the maximum circular velocity of the satellite's dark matter halo ($ v_{\rm max}$; red) and the total orbital angular momentum ($j$; purple). \textbf{Right:} The estimated mass of M31 using the \textit{HST}+\textit{Gaia} M31 $v_{\rm tan}$ zero-point. The latter results are significantly higher since the 3D velocity of M33 relative to M31 increased between the old (\textit{HST}+sats) vs. new (\textit{HST}+\textit{Gaia}) M31 PM measurements (see Table \ref{tab:obsdata}). Overall, IllustrisTNG-Dark results are systematically lower by 5\% as compared to the same results using the  Illustris-Dark simulation \citepalias{patel17b}.}
    \label{fig:M33_HST_Gaia}
    \end{center}
\end{figure*}

As this work employs IllustrisTNG, we estimate the mass of M31 using only the properties of M33 and the IllustrisTNG-Dark simulation to see how the results compare to those using Illustris-1-Dark. For consistency, we adopt the same values for the observed M33 data as in \citetalias{patel17b}, rather than the revised values in Table \ref{tab:obsdata}.

We choose a prior sample from IllustrisTNG-Dark using identical criteria as in \citetalias{patel17b}. This includes choosing only those host halos that contain a massive satellite analog with $v_{\rm max} > 70 \rm \,km\, s^{-1}$, that resides within the virial radius of its host galaxy at $z\approx0$, and that has a minimal subhalo mass of $10^{10}\,M_{\odot}$ at $z\approx0$. We build the prior sample statistics by choosing all halo systems with massive satellite analogs satisfying these properties from snapshots corresponding to $z=0-0.26$ (or snapshots 80-99 in IllustrisTNG-Dark). This yields 24,964 halo systems that constitute the prior sample. The original prior sample in \citetalias{patel17b} contained 19,653 halos. 

Using this prior sample, we calculate likelihood functions as in Eqs. 7-10 from \citetalias{patel17b} and marginalize over \Mvir using importance sampling. For the same exact M33 properties adopted in \citetalias{patel17b} (including the original angular momentum value), we find an M31 mass of \Mvir$=1.30^{+1.40}_{-0.66}\times10^{12}\, M_{\odot}$. These results are illustrated in Figure \ref{fig:M33_HST_Gaia}. In \citetalias{patel17b}, we reported an M31 mass of \Mvir$=1.37^{+1.39}_{-0.75}\times10^{12}\, M_{\odot}$ for the momentum method. The IllustrisTNG-Dark results are systematically lower by $\sim$5\% compared to the Illustris-Dark results \citepalias{patel17b}, however, they are consistent within the corresponding credible intervals. A preliminary analysis of M31 stellar mass analogs, chosen as the primary halos whose corresponding stellar masses via the \citet{moster13} abundance matching relationship are in the range $5-10 \times 10^{10}\, M_{\odot}$, in both Illustris-Dark and IllustrisTNG-Dark show that this could be an artifact of a systematic position offset between satellites located at $<$ 400 kpc with a 1:10-1:100 mass ratio with these M31 analogs. This exercise shows the median position offset between satellites and their hosts in IllustrisTNG-Dark is also $\sim$5\% lower than their counterparts in Illustris-Dark. These offsets, their origins, and implications will be discussed in detail in Chamberlain et al., in preparation. 

Since both M31 and M33 have a new set of independent PM measurements based on \textit{Gaia} DR2 data, and updated distances, we also compute the estimated mass of M31 using the combination of \textit{HST}+\textit{Gaia} M31 phase space information and the \textit{VLBA}+\textit{Gaia} M33 phase space information. The corresponding observed properties are listed in Table \ref{tab:obsdata} and the resulting M31 mass is\Mvir$=1.87^{+2.20}_{-0.79}\times10^{12}\, M_{\odot}$. These results are illustrated in the right panel of Figure \ref{fig:M33_HST_Gaia}. Given the increase in M33's relative position and velocity with the \textit{HST}+\textit{Gaia} M31 v$_{\rm tan}$ zero-point, it is unsurprising that these results also increase, as illustrated by comparing the purple curves in the left and right panels in Figure \ref{fig:M33_HST_Gaia}.

\section{Derivation of Importance Weights for Multiple Satellites}
\label{app:derivation}

The joint posterior distribution of a halo's log mass $m \equiv \log_{10} M_{\rm vir}$ and the latent properties $\bm{X} = \{\bm{x}_1,\ldots,\bm{x}_{N_{\rm sat}}\}$ of its $N_{\rm sat}$ subhalos, conditional on the measurements $\bm{D} = \{\bm{d}_1,\ldots,\bm{d}_{N_{\rm sat}}\}$, is given by Bayes' Theorem (Eq. \ref{eq:bayes1}):
\begin{equation}
    P(m, \bm{X} |\, \bm{D}) \propto P( \bm{D}|\, m, \bm{X} \}) \times P(m, \bm{X}| \, \bm{C})
\end{equation}
We invoke the reasonable assumption that, conditional on the true latent properties, the probability distribution of the measurements has no additional dependence on $m$. This implies that the measurement errors are independent of $m$. Furthermore, we assume that the measurements of each satellite's properties, conditional on the true latent values of those properties, are mutually independent of the other satellites' properties and their measurements. These reasonable assumptions allow us to write the second term as:
\begin{equation}
 P( \{\bm{d}_1,\ldots,\bm{d}_{N_{\rm sat}} \}|\, m, \{\bm{x}_1,\ldots,\bm{x}_{N_{\rm sat}} \}) = \prod_{s=1}^{N_{\rm sat}} P(\bm{d}_s |\, \bm{x}_s).
\end{equation}
Expectations of functions of the log mass, $f(m)$, with respect to the posterior distribution (Eq. \ref{eq:posterior_1}) can be written as:
\begin{equation}
    \mathbb{E}[f(m) | \, \bm{D}] = \frac{\int f(m) \left[ \prod_{s=1}^{N_{\rm sat}} P(\bm{d}_s |\, \bm{x}_s) \right] \times P(m, \{\bm{x}_1,\ldots,\bm{x}_{N_{\rm sat}} \}| \, \bm{C}) \, {\rm d}m \, {\rm d}x_1,\ldots,{\rm d}x_{N_{\rm sat}}}{\int \left[ \prod_{s=1}^{N_{\rm sat}} P(\bm{d}_s |\, \bm{x}_s) \right] \times P(m, \{\bm{x}_1,\ldots,\bm{x}_{N_{\rm sat}} \}| \, \bm{C}) \, {\rm d}m \, {\rm d}x_1,\ldots,{\rm d}x_{N_{\rm sat}}},
\end{equation}
where the denominator is the normalization term. To derive the self-normalized importance weights, we approximate both integrals as Monte Carlo sums over $n$ independent draws from the prior, i.e. the halo systems from the simulation, indexed as $j = 1,\ldots,n$,
\begin{equation}
m^j, \{\bm{x}_1^j,\ldots,\bm{x}^j_{N_{\rm sat}} \} \sim P(m, \{\bm{x}_1,\ldots,\bm{x}_{N_{\rm sat}} \}| \, \bm{C}).
\end{equation}
The posterior expectation is estimated from these samples as:
\begin{equation}
    \mathbb{E}[f(m) | \, \bm{D}] \approx \frac{\sum_{j=1}^n f(m^j) \prod_{s=1}^{N_{\rm sat}} P(\bm{d}_s |\, \bm{x}^j_s)}{\sum_{j=1}^n \prod_{s=1}^{N_{\rm sat}} P(\bm{d}_s |\, \bm{x}^j_s)} 
    = \sum_{j=1}^n f(m^j) \, w_j
\end{equation}
where
\begin{equation}
    w_j \equiv \frac{\prod_{s=1}^{N_{\rm sat}} P(\bm{d}_s |\, \bm{x}^j_s)}{\sum_{i=1}^n \prod_{s=1}^{N_{\rm sat}} P(\bm{d}_s |\, \bm{x}^i_s)}
\end{equation}
are the self-normalized importance weights, as in Eq. \ref{eq:weights_nsat}.  Note that $\sum_{j=1}^n w_j = 1$.

\end{document}